\documentclass[]{aastex631}

\begin{document}

\title{Redox evolution of the crystallizing terrestrial magma ocean and its
influence on atmosphere outgassing}

\correspondingauthor{Maxime Maurice}
\email{maximemaurice@protonmail.com}

\author[0000-0001-8804-120X]{Maxime Maurice}
\affiliation{Rice University \\
Earth, Environmental and Planetary Science Department \\
6100 Main Street, MS 126, Houston, TX 77005, USA}

\author[0000-0001-5392-415X]{Rajdeep Dasgupta}
\affiliation{Rice University \\
Earth, Environmental and Planetary Science Department \\
6100 Main Street, MS 126, Houston, TX 77005, USA}

\author[0000-0001-9425-8085]{Pedram Hassanzadeh}
\affiliation{Rice University \\
Earth, Environmental and Planetary Science Department \\
6100 Main Street, MS 126, Houston, TX 77005, USA}
\affiliation{Rice University \\
Mechanical Engineering Department \\
100 Main Street, MS 126, Houston, TX 77005, USA}



\begin{abstract}

Magma oceans are episodes of large-scale melting of the mantle of terrestrial planets. The energy delivered by the Moon-forming impact induced a deep magma ocean on the young Earth, corresponding to the last episode of core-mantle equilibration. The crystallization of this magma ocean led to the outgassing of volatiles initially present in the Earth's mantle, resulting in the formation of a secondary atmosphere. During outgassing, the magma ocean acts as a chemical buffer for the atmosphere via the oxygen fugacity, set by the equilibrium between ferrous- and ferric-iron oxides in the silicate melts. By tracking the evolution of the oxygen fugacity during magma ocean solidification, we model the evolving composition of a C-O-H atmosphere. We use the atmosphere composition to calculate its thermal structure and radiative flux. This allows us to calculate the lifetime of the terrestrial magma ocean. We find that, upon crystallizing, the magma ocean evolves from a mildly reducing to a highly oxidized redox state, thereby transiting from a CO- and H$_2$-dominated atmosphere to a CO$_2$- and H$_2$O-dominated one. We find the overall duration of the magma ocean crystallization to depend mostly on the bulk H content of the mantle, and to remain below 1.5 millions years for up to 9 Earth's water oceans' worth of H. Our model also suggests that reduced atmospheres emit lower infrared radiation than oxidized ones, despite of the lower greenhouse effect of reduced species, resulting in a longer magma ocean lifetime in the former case. Although developed for 
a deep magma ocean on Earth, the framework applies to all terrestrial planet and exoplanet magma oceans, depending on their volatile budgets.

\end{abstract}



\section{Introduction}

During their formation, terrestrial planets are thought to go through phases of large-scale melting of their silicate mantles referred to as ``magma oceans'' (MO) \citep{ElkinsTanton2012}. Such MOs would represent the most important episodes of rapid chemical equilibration between the mantle and the atmosphere in a planet's history \citep{AbeMatsui1985}. Conversely, the solid mantle is chemically isolated, and chemical exchanges with the atmosphere proceed through extremely sluggish processes (for instance volcanism or subduction of volatile-rich oceanic crust). Therefore, outgassing taking place during MO solidification largely sets the initial stage for the planetary atmospheres evolution. Characterizing it properly is thus key to understanding the differences between terrestrial planets in the solar system. Understanding MO-equilibrated atmosphere is also crucial to interpreting upcoming observations of exoplanetary atmospheres.

Terrestrial MO atmospheres, over the temperature range considered in this study (1500 K - 3000 K), are thought to be largely dominated by the most abundant volatile elements (e.g. H and C). In addition to being of prime interest as life-essential elements, volatile elements also form gaseous species with strong absorption features in the infrared, affecting the radiative state of the atmosphere. They are, therefore, the focus of many MO-atmosphere fractionation studies \citep{KuramotoMatsui1996,Dasgupta2013,Hirschmann2016,DasguptaGrewal2019}.

Early work by \citet{AbeMatsui1985,AbeMatsui1988} laid the ground work for research on the link between the terrestrial MO and an outgassed H$_2$O-CO$_2$ atmosphere, delivered by volatiles from accreting planetesimals. They pointed out that the thermal blanketing effect of a thick steam and ${\rm CO}_2$ atmosphere prevents efficient cooling of the MO. Indeed, the presence of greenhouse gases in the atmosphere reduces the thermal radiative heat flux from the MO surface, while leaving the high wavelength instellation largely unaltered, thereby reducing the net cooling of the planet and slowing down its solidification. Later studies by \citet{ElkinsTanton2008} and \citet{Lebrun2013}, considering the same gaseous species, emphasized the difference in the outgassing patterns of the two gases: CO$_2$, having a low solubility in silicate melts, is readily outgassed from the MO since the beginning, while water, being much more soluble, remains largely dissolved until the late stages of crystallization. Addressing the thermal evolution of the MO, these studies showed that the initial volatile budget, especially that of water, has a first order influence on the MO lifetime, whereby more volatiles induce a longer crystallization. The MO lifetime can even become infinite when the incoming stellar flux exceeds a threshold, associated with the buffering effect of water condensation on the atmospheric infrared emission \citep{Hamano2013,Salvador2017}, above which the MO no longer cools. This scenario might have been applicable to Venus, in which case only the dessication of its steam atmosphere by H$_2$ escape could have ended the magma ocean phase. More recently, \citet{Bower2019} pointed out a mechanism by which the late outgassing of water decreases the partial pressure of CO$_2$, casting a more complex picture of the outgassing pattern than observed thus far. In a two related studies, \citet{Nikolaou2019} and \citet{Katyal2019} also investigated the sensitivity of the terrestrial MO duration to a series of parameters including the initial volatile budget, the convective regime in the MO, and the parametization of the radiative flux through the atmosphere, confirming the prime influence of H budget.

All of these studies focused on oxidized atmospheres (composed of ${\rm H}_2{\rm O}$ and ${\rm CO}_2$). Lately, a significant effort has been undertaken to broaden the scope to more general chemistry. Estimating the MO's oxygen fugacity ($f_{{\rm O}_2}$) based on novel experiments on molten peridotite, \citet{Sossi2020} suggested that the early terrestrial atmosphere inherited from MO outgassing might have been akin to today's Venus atmosphere, i.e. dominated by CO$_2$ and N$_2$. Bringing S into the picture while treating $f_{{\rm O}_2}$ as a free parameter, \citet{Gaillard2022b} showed that reduced atmospheres are dominated by H-C species while oxidized ones are dominated by C-N-S species. They showed in particular that the present-day surficial reservoirs of C and N on the Earth can be matched in the atmosphere at equilibrium with a deep MO, inferring that later processes might have played only minor roles in the surface inventory of these elements. Other studies focused on quantifying the radiative state of the outgassed atmosphere, beyond CO$_2$ and H$_2$O. For example \citet{Lichtenberg2021} analyzed the effect of different gaseous species individually (in single-gas model atmospheres). They distinguished between three groups of volatiles: ${\rm O}_2$, ${\rm N}_2$ and ${\rm CO}$ have both a low solubility and poor greenhouse effect, yielding immediately outgassed atmospheres and short-lived MOs. ${\rm CO}_2$, ${\rm CH}_4$ and ${\rm H}_2{\rm O}$ have stronger greenhouse effect and various solubilities (generally higher), inducing longer MO and evolving atmospheric pressure. Finally ${\rm H}_2$, was found to have the strongest effect on the lifetime of the MO, due to its strong collision-induced absorption at high pressure. Focusing on the H- and C-bearing systems (i.e. H$_2$O, H$_2$, CO$_2$ and CO), \citet{Katyal2020} calculated the redox speciation in the atmosphere overlying different stages of the MO evolution computed by \citet{Nikolaou2019}. By varying the $f_{{\rm O}_2}$ of the MO at equilibrium with the atmosphere as a free parameter, they found that the transition from reduced to oxidized conditions translates in a decrease of the radiative cooling flux. Taking one step further, \citet{Bower2022} computed the chemical evolution of an atmosphere outgassed throughout the crystallization of a MO with an oxygen fugacity at 0.5 log-unit below the iron-w\"ustite buffer (following \citet{Sossi2020}). By investigating the bulk C/H ratio as well as the volatile budget, they found a general evolution towards a more oxidized atmosphere as crystallization proceeds. Furthermore, the combined outgassing of H- and C-bearing species generally results in a transition from a CO-dominated early atmosphere to an H$_2$O-dominated late one, although a major part of the H can remain trapped in interstitial melts in the mantle.

These previous studies, therefore, established that MO oxidation state is a crucial parameter governing the chemistry of atmospheres. As long as the MO remains in equilibrium with the atmosphere and dominates it in terms of mass, one can assume that the MO chemistry buffers the oxidation state of the atmosphere. The redox chemistry in the MO is initially set by the equilibrium between metallic iron and iron oxide during the formation of the core. For an Earth-like composition, this event buffered the $f_{{\rm O}_2}$ to two log-units below the iron-w\"ustite buffer (setting $\Delta{\rm IW}=-2$, where $\Delta{\rm IW}$ is the deviation between the actual $f_{{\rm O}_2}$ of the MO and that corresponding to the iron-w\"ustite buffer). Subsequently, after metallic iron segregates to the core, the redox state of the MO is governed by the equilibrium between ferrous- (${\rm Fe}^{2+}$) and ferric- (${\rm Fe}^{3+}$) iron oxides (i.e. ${\rm FeO}$ and ${\rm FeO}_{1.5}$). \citet{Hirschmann2012} proposed that the pressure-dependence of this equilibrium should induce a negative gradient in the $f_{{\rm O}_2}$ versus depth profile in a MO where the ratio between the ${\rm FeO}$ and ${\rm FeO}_{1.5}$ concentrations is homogenized by convective mixing. In this case, depending on the effective pressure of core-mantle equilibrium, the initial $f_{{\rm O}_2}$ at the surface (where the MO equilibrates with the atmosphere) may deviate from $\Delta{\rm IW}=-2$. Recent studies using experimental \citep{Zhang2017,Armstrong2019} or first principle molecular dynamics \citep{Deng2020} methods have confirmed it. For a constant ferric-to-total iron oxide ratio (${\rm Fe}^{3+}/\Sigma {\rm Fe}$) throughout the MO, the oxygen fugacity increases (relative to the iron-w\"ustite buffer) with decreasing pressure (except in the top $\sim10$ GPa). On Earth, the last major event of core-mantle equilibration is thought to be associated with the Moon-forming impact, which produced a deep MO. It is thus possible that the surface $f_{{\rm O}_2}$ was higher than IW-2 \citep{Wood2006}.

Previous studies on MO-atmosphere interactions have considered these uncertainties on the redox state by varying the $f_{{\rm O}_2}$ as a free parameter \citep{Katyal2020,Sossi2020,Gaillard2022b,Bower2022}. However, the quantity ${\rm Fe}^{3+}/\Sigma {\rm Fe}$ which governs the MO $f_{{\rm O}_2}$, is expected to evolve as the MO crystallizes, because ferric iron is more incompatible than ferrous iron in silicate minerals \citep{Boujibar2016,DavisCottrell2021,RudraHirschmann2022}. Therefore, ${\rm Fe}^{3+}/\Sigma {\rm Fe}$ is expected to increase, making the MO more oxidized as it crystallizes. In other words, the redox evolution is expected to be coupled to the thermal evolution of the MO. In the present study, we self-consistently compute the evolution of the oxygen fugacity in the MO by tracking ${\rm Fe}^{3+}/\Sigma {\rm Fe}$. To do this, we use the parametrization linking $f_{{\rm O}_2}$ and ${\rm Fe}^{3+}/\Sigma {\rm Fe}$ obtained by \citet{Hirschmann2022}, fitted on a $P$-$T$-composition space encompassing high pressures (from \citet{Deng2020}), high (MO-like) temperatures, and peridotitic composition relevant for the MO (from \citet{Sossi2020}). Based on the surface $f_{{\rm O}_2}$ of the MO, we compute the redox speciation in an H-C-O atmosphere. From the composition of the outgassed atmosphere, we calculate the cooling flux of the MO using a convective-radiative model allowing us to capture influential effects on the thermal evolution of the coupled atmosphere and MO system, which were not accessible to simpler atmosphere models. We vary the crystallization regime (i.e. the efficiency of crystals settling vs entrainment), the bulk volatile budget, and the bulk C/H ratio, to investigate their influences on the composition of the outgassed atmosphere, the thermal evolution and lifetime of the MO crystallization, and the retention of volatile in the mantle. We apply this approach to the case of the post Moon-forming impact terrestrial MO.

\section{Model}
\label{sec:model}
This study has three main model components. First, the crystallization model which, depending on the dynamical regime of settling of the crystallizing minerals, allows us to relate the potential temperature of the MO to its depth, and to calculate elemental fractionation between solids and liquids. Second, the redox chemistry model which, based on the evolution of the ${\rm Fe}^{3+}/\Sigma{\rm Fe}$ ratio in the MO allows us to calculate the $f_{{\rm O}_2}$ and, in turn, the chemical speciation of the atmosphere in equilibrium with the MO. Considering mass conservation of volatile elements as well as solubility equilibria, we track the outgassing of the MO and the formation of the atmosphere.  The outputs of the chemical model are the partial pressures of each species present in the atmosphere and their concentration in the MO (as well as $f_{{\rm O}_2}$ and ${\rm Fe}^{3+}/\Sigma{\rm Fe}$). Third, the thermal evolution model, which allows us to calculate the cooling flux of the MO radiating through its outgassed atmosphere, and thus constrain its lifetime.

In the following, Section \ref{subsec:MO_crystallization} presents the technical definition of the MO depending on the dynamical regime of crystals settling. Section \ref{subsec:chemical_model} presents the redox chemical model for both iron oxides and volatile species, Finally, Section \ref{subsec:thermal_model} introduces the thermal evolution model.

\subsection{Magma ocean crystallization}
\label{subsec:MO_crystallization}
The crystallization of the terrestrial MO is generally thought to proceed from the bottom up to the surface as a consequence of the stronger pressure gradient of the melting curves compared to the temperature profile in the MO \citep{Solomatov2015}, and/or of the higher density of solid crystals. Limits to the validity of both of these assumptions \citep{Stixrude2009,Caracas2019} have been pointed out, which could lead to a middle-out crystallization of the MO, and possibly to the formation of a basal MO \citep{Labrosse2007}. In this work, we restrict the study to a terrestrial MO 55 GPa-equivalent in depth \citep{Deng2020}, as it represents the effective pressure of core-mantle equilibration. It is likely too shallow for a middle-out crystallization pathway to apply.

Under the assumption that crystals efficiently settle at the bottom of the MO and form a compact cumulates pile, the distinction between the MO and its solid cumulates is straightforward. However, when crystals remain in suspension and a partially molten domain exists, this distinction becomes blurred. Here, we envisage both cases, and refer to them as ``fractional'' and ``equilibrium'' crystallization, respectively. In the latter case, the ``MO'' refers to the layer in a liquid-like dynamical regime (undergoing turbulent convection), while ``solid cumulates'' refers to the layer in a solid-like dynamical regime (undergoing creep deformation). The bottom of the MO is then defined as the depth at which this transition between two dynamical regimes occurs, as described in the next Section.

The fate of crystals is thus essential to the distinction between fractional and equilibrium crystallization. The settling or entrainment of the crystallizing minerals in an MO is the result of a competition between convective currents and buoyancy forces \citep{SolomatovStevenson1993a,Patocka2020,Patocka2022}. It is poorly constrained, mainly due to the difficulties in reproducing MO-like dynamical regimes either experimentally or numerically.

Notations in use in Section \ref{subsec:MO_crystallization} are listed in Table \ref{tab:param1}.
\begin{figure}
    \centering
    \includegraphics[width=\textwidth]{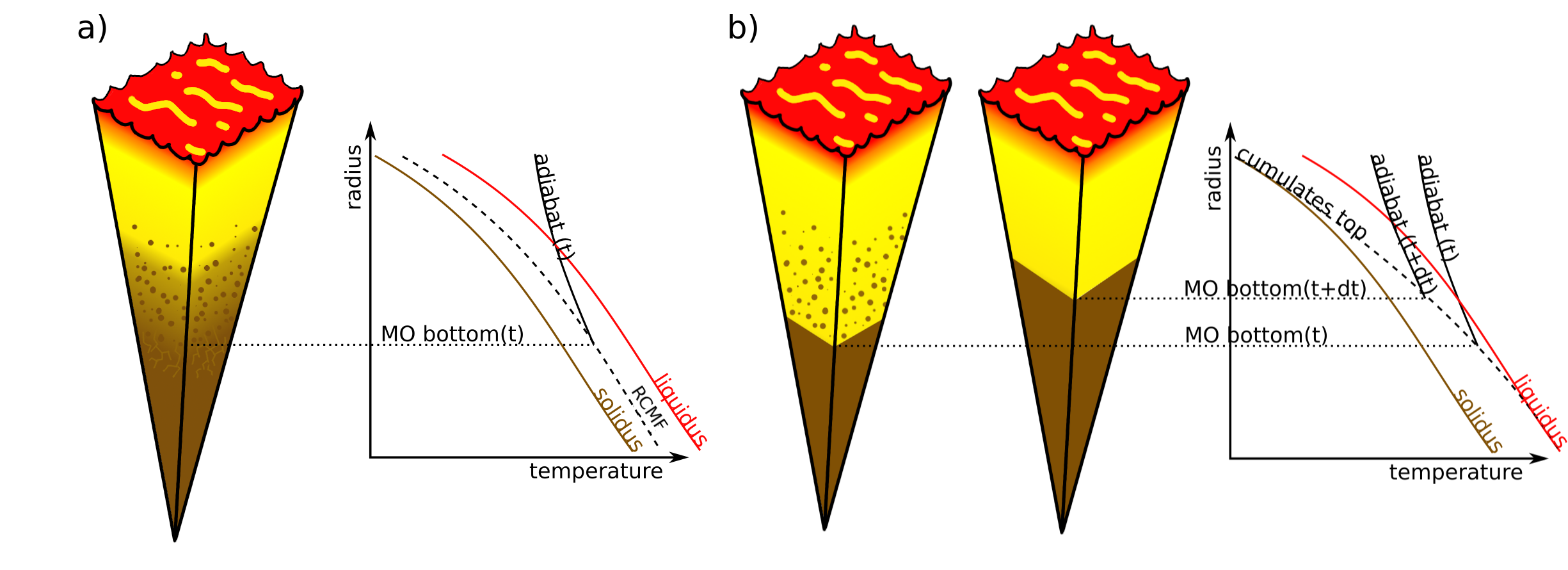}
    \caption{Two different crystallization scenarios: a) equilibrium crystallization where crystals remained entrained by the flow and the bottom of the MO is where the melt fraction reaches the RCMF, and b) fractional crystallization where crystals settle at the bottom and form a compacted cumulates pile, whose top corresponds to the bottom of the MO. Curves are simply illustrative and do not correspond to the actual melting curves and MO adiabat.}
    \label{fig:crystallization_scenarios}
\end{figure}

\subsubsection{Equilibrium crystallization}
\label{subsubsec:equilibrium_crystallization}
In this scenario (represented in Figure \ref{fig:crystallization_scenarios}a), we assume that convective currents in the MO are stronger than the buoyancy force acting on the crystallizing solids. Thus crystals remain entrained and the solidifying medium has a melt fraction parametrized as a function of pressure, temperature, and the melting curves:
\begin{equation}
    \phi(P)=\frac{T(P)-T_{\rm sol}(P)}{T_{\rm liq}(P)-T_{\rm sol}(P)},
    \label{eq:melt_fraction}
\end{equation}
where $P$ is the ambient (lithostatic) pressure, $T$ the corresponding temperature on the MO adiabat (see Section \ref{subsubsec:adiabat}), and $T_{\rm sol}$ and $T_{\rm liq}$ the corresponding solidus and liquidus temperatures, respectively \citep{Fiquet2010}. The medium behaves dynamically as a liquid until the ambient melt fraction reaches a threshold value, referred to as the rheologically critical melt fraction (RCMF) \citep{SolomatovStevenson1993b}. When the melt fraction crosses the RCMF, solid crystals start to form an interconnected lattice, which supports the deformation, resulting in a drastic increase of the effective viscosity \citep{Costa2009}. The value of the RCMF varies depending on multiple parameters like the shape and size of the crystals. In line with previous studies, we use a value of 0.4 \citep{Solomatov2015,Monteux2016,Nikolaou2019} for the main part of this study, but we also vary it in Section \ref{subsec:volatile_mantle}. The MO extends from the surface down to a melt fraction of 0.4, and always contains a significant mass of free-floating crystals. Similarly, the solidifying cumulates retain a significant quantity of trapped melt, that is assumed to freeze in place as the solid mantle further cools down, eventually contributing to the mantle's volatile budget as discussed in Section \ref{subsec:volatile_mantle}. The crystals in suspension in the MO remain in chemical equilibrium with the melt, while the melt trapped in the solid cumulates is chemically isolated from the MO and its composition is no longer affected by MO fractionation. The initial temperature profile of the equilibrium crystallization scenario is represented in Figure \ref{fig:initial_redox_profiles}b.

\subsubsection{Fractional crystallization}
\label{subsubsec:fractional_crystallization}
In this scenario (represented in Figure \ref{fig:crystallization_scenarios}b), we assume that the buoyancy force acting on the crystallizing phases is stronger than the convective currents in the MO. Thus, the crystals settle at the bottom of the MO and form a compact cumulates pile from which melt is entirely extracted. The mass of solid phases formed over one crystallization step is used to compute an equivalent-volume layer thickness at the top of the current cumulate pile and update the depth of the MO accordingly.

We assume that, at the onset of the MO's crystallization, its bottom is at the liquidus temperature at a pressure of 55 GPa. Thus the fractional crystallization cases exhibits a higher initial potential temperature (2939 K) than equilibrium crystallization ones (2571 K), where the bottom of the MO is at the RCMF at 55 GPa.  The initial temperature profile of the fractional crystallization scenario is represented in Figure \ref{fig:initial_redox_profiles}b.
The minerals are in chemical equilibrium with the MO only during the crystallization step they are formed in, and then become chemically isolated from the next step on.

\subsubsection{MO adiabat}
\label{subsubsec:adiabat}
The temperature profile in the MO is used to calculate $T$-dependent quantities, in particular the melt fraction (Equation \ref{eq:melt_fraction}), but also the equations of state (see below) or the MO viscosity (Equation \ref{eq:VFT}). The temperature follows an adiabat \citep{Solomatov2015}:
\begin{equation}
    \frac{\partial T}{\partial P}=\frac{\alpha T}{\rho c_p},
    \label{eq:MO_adiabat}
\end{equation}
where $\alpha$, $\rho$ and $c_p$ are the thermal expansivity, density and heat capacity of silicate melt, respectively. They follow the equation of state for molten MgSiO$_3$ from \citet{deKokerStixrude2013} (calculated using burnman \citep{burnman}). Because the equations of state depend on $T$, Equation \ref{eq:MO_adiabat} is a non-linear ordinary differential equation, that we solve given a boundary condition using a Runge-Kutta 4$^{\rm th}$ order stepping method. The potential temperature $T_{\rm pot}$ is the boundary condition at the MO surface (although different  from the surface temperature due to the thermal boundary layer driving convection in the MO, see Section \ref{subsubsec:convective_flux}) from which the adiabat is integrated. Thus, $T_{\rm pot}$ characterizes entirely the temperature profile, and in turn, the depth of the MO, and all quantities depending on it. These quantities can be conceived as state functions of the state variable $T_{\rm pot}$.

\begin{table}[h!]
    \centering
    \begin{tabular}{l l l l}
        {\bf Symbol} & {\bf Quantity} & {\bf Value/Expression} & {\bf Units} \\ \hline
        $T$ & temperature & variable & K \\
        $T_{\rm liq}$ & liquidus temperature & \citet{Fiquet2010} & K \\
        $T_{\rm sol}$ & solidus temperature & \citet{Fiquet2010} &K \\
        $P$ & lithostatic pressure & variable & Pa \\
        $\phi$ & melt fraction & Eq. \ref{eq:melt_fraction} & - \\
        $\alpha$ & thermal expansivity & \citet{deKokerStixrude2013} & K$^{-1}$ \\
        $\rho$ & density & \citet{deKokerStixrude2013} & kg/m$^3$ \\
        $c_p$ & specific heat capacity of the mantle & \citet{deKokerStixrude2013} & J/kg \\
        \hline
    \end{tabular}
    \caption{Notation used in Section \ref{subsec:MO_crystallization}. Notice that in tables \ref{tab:param1}, \ref{tab:param2} and \ref{tab:param3},``variable'' means spatially changing while ``evolving'' means temporally changing (when both apply, only the first one is indicated).}
    \label{tab:param1}
\end{table}

\subsection{Chemical model}
\label{subsec:chemical_model}
The chemical model relies on three considerations: 1) mass conservation of Fe$^{2+}$, Fe$^{3+}$, H and C, 2) redox equilibria between FeO (i.e. Fe$^{2+}$) and FeO$_{1.5}$ (i.e. Fe$^{3+}$), H$_2$ and H$_2$O, and CO and CO$_2$ respectively, and 3) dissolution equilibria: crystal-melt partitioning of Fe$^{2+}$ and Fe$^{3+}$, and gas-melt dissolution of H$_2$ and H$_2$O, and CO and CO$_2$. Notice that we consider both H and C perfectly incompatible.
Redox equilibrium of the FeO-FeO$_{1.5}$ system are calculated using the parametrization from \citet{Hirschmann2022} (Equation \ref{eq:fO2_Hirschmann2022}), and redox equilibria between the H-bearing and C-bearing species are calculated using temperature-dependant reaction constants from JANAF tables \citep{JANAF} (Section \ref{apx:volatile_redox}) as well as the calculated $f_{{\rm O}_2}$ for. Crystal-melt partitioning of FeO and FeO$_{1.5}$ are calculated using a set of partition coefficients (Section \ref{apx:part_coefs}), and gas-melt equilibrium is calculated using solubility laws (Equation \ref{eq:Henry_law}, Table \ref{tab:solubility_coefs}).

The pathway of the chemical model is as follows: from Fe$^{2+}$ and Fe$^{3+}$ mass conservation, we track the relative abundances (assumed homogeneous in the MO) of both iron oxides, expressed as the molar ferric-to-total iron ratio, ${\rm Fe}^{3+}/\Sigma {\rm Fe}$. From ${\rm Fe}^{3+}/\Sigma {\rm Fe}$, we calculate the $f_{{\rm O}_2}$ using the redox equilibrium parametrization. We then use $f_{{\rm O}_2}$ to calculate the relative abundances of the gaseous species in the atmosphere. Finally, the system is closed by considering dissolution equilibria of volatile species and mass conservation of C and H.

Notations in use in Section \ref{subsec:chemical_model} are listed in Table \ref{tab:param2}.

\subsubsection{Mass conservation}
\label{subsubsec:mass_cons}
Mass conservation is enforced on a system consisting of the crystallizing MO (as defined in either Section \ref{subsubsec:equilibrium_crystallization} or \ref{subsubsec:fractional_crystallization}) and the atmosphere. Notice that, for the sake of generality, here the mass conservation model is described for a hypothetical element that can enter solid crystals, melt, and atmosphere, while in our case Fe$^{2+}$ and Fe$^{2+}$ are distributed between crystals and melt while C and H are distributed between melt and atmosphere only. 
At each crystallization step, the inventory of element (or ion for Fe) $e$ is distributed between 3 reservoirs: entrained crystals in the MO, melt in the MO, and atmosphere:
\begin{equation}
    M_{\rm syst}^e=M_{\rm crys}X_{\rm crys}^e+M_{\rm liq}X_{\rm liq}^e+M_{\rm atm}X_{\rm atm}^e,
    \label{eq:mass_distrib}
\end{equation}
where $M_{\rm syst}^e$ is the total mass of $e$ in the system (MO + atmosphere), $M_{\rm crys}$, $M_{\rm liq}$ and $M_{\rm atm}$ the masses of crystals in the MO, liquid in the MO, and gas in the atmosphere, respectively, and $X_{\rm crys}^e$, $X_{\rm liq}^e$ and $X_{\rm atm}^e$ the mass fractions of $e$ in the corresponding reservoirs. Notice that in the fractional crystallization scenario, $M_{\rm liq}$ corresponds to the mass of the MO and $M_{\rm crys}=0$.

The MO, considered as an open system, loses a mass $dM_{\rm syst}$ to the cumulates pile during one crystallization step (see Appendix \ref{apx:mass_cons}). We neglect atmospheric loss, which would be another mass sink for the system in the case of volatiles; however we discuss its potential effects in Section \ref{subsec:atm_escape}. $dM_{\rm syst}$ is composed of a fraction $\phi_{\rm RCMF}$ of liquid and a fraction $(1-\phi_{\rm RCMF})$ of crystals (notice that this also applies to the fractional crystallization scenario by considering $\phi_{\rm RCMF}=0$ in this case). Accordingly, the mass of each element $e$ in the system decreases by:
\begin{equation}
    dM_{\rm syst}^e=\left(\phi_{\rm RCMF}X_{\rm liq}^e+(1-\phi_{\rm RCMF})X_{\rm crys}^e\right)dM_{\rm syst},
    \label{eq:dM_x}
\end{equation}
Notice that in the fractional crystallization case, $X_{\rm crys}^e$ represents the content $e$  in the crystals that form and settle during the current crystallization step, and can be determined in the same way as for equilibrium crystallization (namely Equation \ref{eq:part_coef} below).

\subsubsection{$f_{{\rm O}_2}$ in the magma ocean}
\label{subsubsec:fO2_in_MO}
During  the core formation event associated with mantle melting and MO creation, equilibrium between Fe-rich metallic melt and Fe-bearing silicate melt (Reaction \ref{eq:iron-wuestite_equilibrium}) sets the $f_{{\rm O}_2}$ of the metal-bearing MO to $\Delta {\rm IW}=-2$ \citep{Wood2006,Rubie2011}, where $\Delta {\rm IW}$ stands for the difference between the actual $f_{{\rm O}_2}$ and that set by the iron-w\"ustite buffer, in log-units. The $f_{{\rm O}_2}$ value corresponding to $\Delta {\rm IW}=-2$ is thus used to compute the initial ${\rm Fe}^{3+}/\Sigma{\rm Fe}$ in the MO silicate liquid, from the equilibrium between ferrous FeO and Fe$_2$O$_3$, (Reaction \ref{eq:ferrous-ferric-iron_equilibrium}).
\begin{eqnarray}
    {\rm Fe}+\frac{1}{2}{\rm O}_2&={\rm FeO},\label{eq:iron-wuestite_equilibrium} \\
    {\rm FeO}+\frac{1}{4}{\rm O}_2&={\rm FeO}_{1.5}.\label{eq:ferrous-ferric-iron_equilibrium}
\end{eqnarray}
The equilibrium constant of Reaction \ref{eq:ferrous-ferric-iron_equilibrium} (given by Equation \ref{eq:fO2_Hirschmann2022} \citep{Hirschmann2022}) depends on pressure and temperature (and composition) \citep{Hirschmann2012,Armstrong2019,Deng2020,Hirschmann2022}, so the effective $P$-$T$ conditions of core-mantle equilibration need to be assumed to calculate the Fe$^{3+}/\Sigma$Fe value corresponding to $\Delta {\rm IW}=-2$. We consider two end-member cases: 1) the ''reduced'' case, in which metal-silicate equilibration occurs at the surface of the MO, simulating the scenario of a MO previously well drained of its metallic iron, with a small impactor emulsifying and supplying metallic melt at near-surface conditions. In this case,  $\Delta {\rm IW}=-2$ at the surface correspond to ${\rm Fe}^{3+}/\Sigma$Fe=0.05 for fractional crystallization, and ${\rm Fe}^{3+}/\Sigma$Fe=0.04 for equilibrium crystallization. 2) the ''oxidized'' case, in which we set $\Delta {\rm IW}=-2$ at the base of the MO at 55 GPa \citep{Deng2020}, i.e. where final ponding of metallic melt could have taken place. This yields ${\rm Fe}^{3+}/\Sigma$Fe=0.11 for fractional crystallization, and ${\rm Fe}^{3+}/\Sigma$Fe=0.08 for equilibrium crystallization. These various end-members for core-mantle equilibration conditions and associated redox state are illustrated in Figure \ref{fig:initial_redox_profiles}.

Using the same parametrization (Equation \ref{eq:fO2_Hirschmann2022}), we then reverse the reasoning and compute the subsequent evolution of the $f_{{\rm O}_2}$ profile in the MO from the evolution of ${\rm Fe}^{3+}/\Sigma{\rm Fe}$ (assumed homogeneous in the MO), as the MO crystallizes. To do so, we track the independent fractionation of Fe$^{3+}$ and Fe$^{2+}$ between solid and melt, using Equations \ref{eq:mass_distrib} and \ref{eq:dM_x}, and their partition coefficients definition:
\begin{equation}
    X_{\rm crys}^e=D_eX_{\rm liq}^e,
    \label{eq:part_coef}
\end{equation}
where $e$ is either FeO or FeO$_{1.5}$ and $D_e$ their respective crystal-silicate melt partition coefficient, specified below. The calculation of initial mass fraction of each oxides from initial molar ${\rm Fe}^{3+}/\Sigma{\rm Fe}$ and the BSE total iron oxides mass fraction is given in Section \ref{apx:part_coefs}.

Ferric iron oxide has a partition coefficient dependent on $P$, calculated based on those for the individual mineral-melt partition coefficients of crystallizing minerals (based on the crystallization sequence of the 2000-km-deep MO from \citet{ElkinsTanton2008}, truncated to 55 GPa, see Appendix \ref{apx:part_coefs}).

We consider a partition coefficient value of 1 for ferrous iron oxide. While ferrous iron oxide is incompatible in most minerals forming the crystallization sequence, the increase of the ferric iron mass fraction only yields very high total iron oxide mass fractions ($>20$ wt\%) in the late stages of fractional crystallization, which would be yet amplified by considering ferrous iron oxide incompatible. This is a natural feature of the fractional crystallization case, which leads to maximum partitioning and is meant to be an end-member case rather than a realistic description. In the equilibrium crystallization case, a value of $D_{{\rm Fe}^{2+}}$ close to 1 ($>0.9$) leads to a final iron oxides mass fraction of 12 wt\%, similar to that of evolved basaltic magmas on Earth. Therefore, for the sake of simplicity, we adopt a partition coefficient of unity, thus amplifying the effect of differential fractionation of Fe$^{2+}$ and Fe$^{3+}$. We do not expect, however, the use of a more realistic value for $D_{{\rm Fe}^{2+}}$ to significantly affect our results.

In the fractional crystallization case, where fractionation is much more important, doing so yields an unrealistically high final mass fraction of total iron oxide. We thus set the partition coefficient of FeO to 1, and let the total iron oxide mass fraction increase up to a maximum of 20\% due to the sole effect of FeO$_{1.5}$ fractionation. While this might artificially increase the effect of iron oxides differential partitionning on MO redox evolution, this value is not far from that used in the equilibrium crystallization case, suggesting that it still captures a realistic behaviour of FeO.

Knowing the evolution of the mass fraction of both oxides (and thus of the molar ${\rm Fe}^{3+}/\Sigma{\rm Fe}$, following the inverse pathway of Section \ref{apx:part_coefs}), we use \citet{Hirschmann2022}'s parametrization to calculate the corresponding $f_{{\rm O}_2}$ at the MO potential temperature and surface pressure, i.e. the conditions at which further equilibria considering $f_{{\rm O}_2}$ are assumed to take place.

\subsubsection{Redox speciation in the atmosphere and volatiles solubility}
\label{subsubsec:fractionation}
At the high temperatures met at the surface of a MO ($>1500$ K), methane is only present in negligible amounts. Thus the C-O-H system is governed by the two following redox equilibria in the atmosphere:
\begin{eqnarray}
    2{\rm H}_2+{\rm O}_2&=2{\rm H}_2{\rm O},\label{eq:H_equilibrium}\\
    2{\rm CO}+{\rm O}_2&=2{\rm CO}_2.\label{eq:C_equilibrium}
\end{eqnarray}
In this section, $e$ refers to one of the volatile elements (H or C) while $s$ refers to one of the molecular species (${\rm H}_2{\rm O}$,${\rm H}_2$,${\rm CO}_2$, or ${\rm CO}$; O$_2$ being excluded from this section's considerations unless explicitly stated). The fugacities of the different gaseous species in the atmosphere are related through the value of the reaction constants of equilibria \ref{eq:H_equilibrium} and \ref{eq:C_equilibrium}, and the oxygen fugacity:
\begin{equation}
    \prod_{s} f_{s}^{c_{s}}=K(T),
    \label{eq:chem_eq}
\end{equation}
where $f_{s}$ is the fugacity of species $s$ involved in the reaction (including ${\rm O}_2$, whose fugacity calculation was presented in the previous section, unlike other gases), $c_{s}$ its corresponding stoichiometric coefficient (signed: positive for products and negative for reactants), $K$ the equilibrium constant and $T$ the effective temperature of the equilibrium.
We consider that the effective temperature at which equilibrium is achieved (i.e. the temperature at which $K$ and $f_{{\rm O}_2}$ are evaluated) is the potential temperature of the MO. This has the advantage of making atmosphere speciation a state function of $T_{\rm pot}$, making the comparison between different models easier. This choice is discussed in Section \ref{sec:results}.
We consider the fugacities to be equal to the partial pressures (expressed in bar) and replace $f_{s}$ with $p_{s}$ in the following.

From the partial pressures of the gaseous species we calculate their content dissolved in the melt $X_{\rm liq}^{s}$ (expressed as mass fraction) using species-specific solubility laws taking the general form:
\begin{equation}
    X_{\rm liq}^{s}=\alpha_{s} p_{s}^{\beta_{s}},
    \label{eq:Henry_law}
\end{equation}
where $\alpha_{s}$ and $\beta_{s}$ are the solubility law coefficients of species $s$ (values listed in Table \ref{tab:solubility_coefs}). We consider volatiles species perfectly incompatible (i.e. having a partition coefficient equal to zero), as a result, $X_{\rm sol}^{s}=0$ holds at all time.

The total atmosphere mass is:
\begin{equation}
    M_{\rm atm}=\frac{4\pi R_{p}^2p_{\rm atm}}{g},
    \label{eq:M_atm}
\end{equation}
with $R_{p}$ and $g$ the planetary radius and gravity, respectively, and $p_{\rm atm}$ the total atmospheric pressure. The mass fraction of species $X$ in the atmosphere is expressed as a function of its partial pressure as:
\begin{equation}
    X_{\rm atm}^{s}=\frac{\mu_{s}}{\mu_{\rm atm}}\frac{p_{s}}{p_{\rm atm}},
    \label{eq:X_atm}
\end{equation}
with $\mu_{s}$ the molecular weight of species $s$ and $\mu_{\rm atm}$ the average molecular weight of the atmosphere. Finally, as an element $e$ can be carried by several species $s$ (e.g. ${\rm H}_2{\rm O}$ and ${\rm H}_2$ for H), its mass inventory (Equation \ref{eq:mass_distrib}) can be written as a function of the partial pressures of gaseous species carrying this element, using \ref{eq:Henry_law}, \ref{eq:M_atm} and \ref{eq:X_atm} and summing over all concerned species:
\begin{equation}
    M^e_{\rm syst}=\sum_{s}\left[\left(M_{\rm liq}\alpha_{s}p_{s}^{\beta_{s}}+\frac{4\pi R_{p}^2}{g}p_{s}\frac{\mu_{s}}{\mu_{\rm atm}}\right)\frac{\mu_{e}\lambda_s^e}{\mu_{s}}\right],
    \label{eq:mass_cons}
\end{equation}
where $\mu_{e}$ is the atomic mass of element ${e}$ and $\lambda_s^e$ the number of atoms of ${e}$ in a molecule of ${s}$.

We solve Equation \ref{eq:chem_eq} two times (once for equilibrium \ref{eq:H_equilibrium} and once for equilibrium \ref{eq:C_equilibrium}) and  Equation \ref{eq:mass_cons} two times (once for conservation of H and once for conservation of C), resulting in 4 equations to solve for the partial pressures of the 4 gaseous species (${\rm O}_2$ not included). These equations are solved using the ``hybrid'' root finding method implemented in the ``root'' function of scipy \citep{scipy}. Contrarily to H and C, O is not considered conserved, but the partial pressure of O$_2$ (i.e the $f_{{\rm O}_2}$) is buffered at the value imposed by Equilibrium \ref{eq:ferrous-ferric-iron_equilibrium}. This assumption only breaks down when the masses of the MO and of the atmosphere become comparable, which is largely past the 99 wt\% MO solidification that we use as our final state. The remaining MO's mass is then $2.32\times10^{22}$ kg, i.e. the equivalent of a 4457 bar, pure-CO$_2$ atmosphere, and higher pressure for a mixed atmosphere. This is a factor 2 larger than the final pressure of our most volatile-rich case. 

\begin{table}[h!]
    \centering
    \begin{tabular}{l l l l}
        {\bf Symbol} & {\bf Quantity} & {\bf Value/Expression} & {\bf Units} \\ \hline
        $\phi_{\rm RCMF}$ & rheologically critical melt fraction & 0.4 & - \\
        $M_{\rm syst}^e$ & mass of element $e$ in the MO + atmosphere system & Eq. \ref{eq:mass_cons} & kg \\
        $dM_{\rm syst}^e$& mass increment of element  $e$  & Eq. \ref{eq:dM_x} & kg \\
        $dM_{\rm syst}$ & mass increment$^a$ of the system & Eq. \ref{eq:dMsyst_equi} or \ref{eq:dM_syst_frac}& kg \\
        $M_{\rm crys}$ &  mass of  crystals  in the system & evolving & kg \\
         $M_{\rm liq}$ &  mass of liquids in the system & evolving  & kg \\
        $M_{\rm atm}$ & mass of atmosphere & Eq. \ref{eq:M_atm} & kg \\
        $X_{\rm crys}^{e/s}$ & mass fraction of element $e$ / species $s$ in the crystals & Eq. \ref{eq:part_coef} & - \\
        $X_{\rm liq}^{e/s}$ & mass fraction of element $e$ / species $s$ in the liquids & Eq. \ref{eq:Henry_law} & - \\
        $X_{\rm atm}^{e/s}$ & mass fraction of element $e$ / species $s$ in the atmosphere & Eq. \ref{eq:X_atm} & - \\
        $f_{{\rm O}_2}$ & oxygen fugacity & evolving & -$^{b}$ \\
        $\Delta{\rm IW}$ & $f_{{\rm O}_2}$ deviation from IW & evolving & log-unit \\
        ${\rm Fe}^{3+}/\Sigma {\rm Fe}$ & molar ferric-to-total iron ratio & evolving & - \\
        $f_{s}$ & fugacity of gaseous species $s$ (except ${\rm O}_2$) & evolving  & -$^{b}$ \\
        $c_{s}$ & (signed) stoichiommetric coefficient of species $s$ & Reactions \ref{eq:ferrous-ferric-iron_equilibrium}, \ref{eq:H_equilibrium}, \ref{eq:C_equilibrium} & - \\
        $K$ & reaction constant & variable  & - \\
        $\alpha_{s}$ & solubility law factor for species $s$ & Table \ref{tab:solubility_coefs} & Pa$^{-\beta_{s}}$ \\
        $\beta_{s}$ & solubility law exponent for species $s$ & Table \ref{tab:solubility_coefs} & - \\
        $R_{p}$ & radius at MO surface & 6731000 & m \\ 
        $g$ & gravity acceleration & 9.81 & m/s$^2$ \\
        $p_{\rm atm}$ & total atmospheric pressure & evolving & Pa \\
        $p_{s}$ & partial pressure of species $s$ & evolving  & Pa \\
        $\mu_{\rm atm}$ & average molecular weight of the atmosphere & evolving$^c$  & kg/mol$^{d}$ \\
        $\mu_{s}$ & molecular weight of species ${s}$ & species-specific & kg/mol$^{d}$ \\
        $\mu_{e}$ & atomic mass of element ${e}$ & element-specific  & kg/mol$^{d}$ \\
        $\lambda_s^{e}$ & number of atoms of element $e$ in one molecule of species $s$ & species- and element-specific  & - \\
        \hline
    \end{tabular}
    \caption{Notation in use in Section \ref{subsec:chemical_model}. Notice that elements are generally designated by the letter $e$ and gaseous species by the letter $s$. $^a$: negative as the MO crystallizes. $^b$: fugacities are equal to partial pressures normalized by a reference pressure of 1 bar. $^c$: the average molecular mass in the atmosphere is the sum of the molecular mass of each species weighted by its molar fraction. $^d$: molecular masses are generally given in g/mol and atomic masses in atomic mass unit (au), which convert into each other with a factor unity; here we use SI units and convert both in kg/mol.}
    \label{tab:param2}
\end{table}

\subsection{Thermal evolution model}
\label{subsec:thermal_model}
The thermal evolution model is used to compute the time-evolution of the MO crystallization and associated atmosphere outgassing. It consists of two main parts: heat conservation in the planet (including core, solid mantle and MO), and radiative heat flux through the greenhouse atmosphere, the latter providing a boundary condition to the former. Notations in use in Section \ref{subsec:thermal_model} are listed in Table \ref{tab:param3}.

\subsubsection{Planetary heat conservation}
\label{subsubsec:heat_cons}
The general form of the heat conservation equation is as follows:
\begin{equation}
    \frac{dE_{\rm th}}{dt}=L\frac{dM_{\rm sol}}{dt}+H_{\rm int}-4\pi R_p^2F_p,
    \label{eq:heat_cons_planet}
\end{equation}
where $E_{\rm th}$ is the thermal energy of the planet (the volume-integral of $\rho c_p T$), $L$ the latent heat of crystallization of silicate, $M_{\rm sol}$ the mass of solids in the whole mantle, $H_{\rm int}$ the internal heating, and $F_p$ the heat flow through the planetary surface (defined positive outward). The left-hand-side term of Equation \ref{eq:heat_cons_planet} corresponds to the secular cooling of the planet. The first term on the left-hand-side corresponds to the release of latent heat associated with MO crystallization, the second to the radioactive decay of heat-producing elements, and the third to the heat flow out through the planetary surface.

An accurate parametrization of the secular cooling term involves resolving various complex heat transfer mechanisms within the young Earth which are, for some, not well understood (e.g. the heat flux at the melting interface between magma ocean and solid mantle \citep{Labrosse2018,Agrusta2020}). For the sake of simplicity, here we consider that the thermal energy of the planet decreases linearly with $T_{\rm pot}$ between the initial and final thermal states (see Section \ref{apx:secular_cooling}). The secular cooling term can thus be written: $dE_{\rm sec}/dt=dE_{\rm th}/dT_{\rm pot}\times dT_{\rm pot}/dt$, where $dE_{\rm th}/dT_{\rm pot}$ is a constant (Table \ref{tab:param3}). This approximation amounts to considering that the whole interior of the Earth convects, intially along a single adiabat, and a boundary layer progressively builds up at the CMB, offsetting the adiabat in the core, preventing it from cooling below the present-day estimated temperature \citep{Andrault2016}.
The final thermal state is similar to (if not colder than) the present-day geotherm and hence likely underestimates the temperature in the Earth at the end of MO crystallization. We thus overestimate the secular cooling term, and in turn, predict an upper bound for the MO lifetime.

$M_{\rm sol}$ depends only on $T_{\rm pot}$, so we can apply the chain rule and write $dM_{\rm sol}/dt=dM_{\rm sol}/dT_{\rm pot}\times dT_{\rm pot}/dt$. Contrarily to the secular cooling term, $dM_{\rm sol}/dT_{\rm pot}$ is not constant but varies as an expression of the shape of the melting curves. We do not account for latent heat release associated with iron crystallization in the core, as inner core crystallization most likely did not start before the end of the MO crystallization \citep{Bono2019} (an agreement with our final thermal state).

The internal heating term corresponds to the heat produced by the decay of the long-lived radioactive elements $^{235}$U, $^{238}$U, $^{232}$Th, and $^{40}$K, distributed in the whole mantle, with abundances indicated in Table 4.2 of \citet{TurcotteSchubert}, calculated at 100 Myr after CAIs formation (our estimate for the Moon-forming impact \citep{Maurice2020}). Considering the short lifetime of the MO compared to the half-lives of these isotopes, we neglect their decrease with time due to radioactive decay and consider $H_{\rm int}$ as a constant (Table \ref{tab:param3}). The heat flux through the MO is the focus of the next Section.

Equation \ref{eq:heat_cons_planet} can thus be recast as the evolution of the MO's potential temperature:

\begin{equation}
    \frac{dT_{\rm pot}}{dt}=\frac{H_{\rm int}-4\pi R_p^2F_p}{dE_{\rm th}/dT_{\rm pot}-LdM_{\rm sol}/dT_{\rm pot}}.
    \label{eq:dTpot_dt}
\end{equation}
The denominator of Equation \ref{eq:dTpot_dt} is always positive (since $dE_{\rm th}/dT_{\rm pot}>0$ and $dM_{\rm sol}/dT_{\rm pot}<0$), so the potential temperature decreases with time as long as the heat flux in the MO dominates internal heating.

\subsubsection{Convective heat flux through the MO}
\label{subsubsec:convective_flux}
Following \citet{Lebrun2013}, the heat flux through the planetary surface ($F_p$) is obtained by equating the convective heat flux from the MO ($F_{\rm conv}$) with the radiative heat flux through the atmosphere ($F_{\rm rad}$). Using results from boundary layer theory \citep{Siggia1994}, the former is parametrized as a function of the temperature drop across the thermal boundary layer at the top of the MO \citep{Lebrun2013}:
\begin{equation}
    F_{\rm conv}=0.089\frac{k(T_{\rm pot}-T_{\rm sfc})}{d}{\rm Ra}^{1/3},
    \label{eq:soft_turbulence}
\end{equation}
where $T_{\rm sfc}$ is the temperature at the surface of the MO, $k$ and $d$ the thermal conducitivity in the MO and its depth, respectively, and Ra the Rayleigh number of the MO, defined as:
\begin{equation}
    {\rm Ra}=\frac{\alpha\rho g(T_{\rm pot}-T_{\rm sfc})d^3}{\eta \kappa},
    \label{eq:Rayleigh_number}
\end{equation}
where $\eta$ is the viscosity of the MO, and $\kappa=k/(\rho c_p)$ its thermal diffusivity. Following \citet{Nikolaou2019}, $\eta$ depends on $T$ following a Vogel-Fulcher-Tammann equation, further affected by the melt fraction in the equilibrium crystallization case (see Appendix \ref{ap:MO_visc} for the expression of $\eta$). As $\rho$, $\eta$, $\alpha$ and $c_p$ are not homogeneous in the MO, ${\rm Ra}$ is calculated using average values for those quantities (arithmetically-averaged for $\rho$, $\alpha$ and $c_p$, and harmonically-averaged for $\eta$, following \citet{Nikolaou2019}).

At high Rayleigh numbers, another convection regime is theorized (referred to as the ``hard'' turbulence regime, as opposed to the ``soft'' turbulence regime described by Equation \ref{eq:soft_turbulence}), which follows a different scaling:
\begin{equation}
    F_{\rm conv}=0.22\frac{k(T_{\rm pot}-T_{\rm sfc})}{d}{\rm Ra}^{2/7}{\rm Pr}^{-1/7}\lambda^{-3/7},
    \label{eq:hard_turbulence}
\end{equation}
where ${\rm Pr}=\eta/(\kappa\rho)$ is the Prandtl number, and $\lambda$ the aspect ratio of the mean flow, which is considered unity following \cite{Lebrun2013,Monteux2016,Nikolaou2019}. While we adopt the soft turbulence scaling in most of the study, we discuss the effect of changing to the hard turbulence scaling in Section \ref{subsec:hard_vs_soft}.

The radiative flux $F_{\rm rad}$ is also a function of $T_{\rm sfc}$ (see next section), so we solve $F_{\rm conv}(T_{\rm sfc})=F_{\rm rad}(T_{\rm sfc})-F_{\rm sun}$ (where $F_{\rm sun}$ is the shortwave flux from the Sun) for $T_{\rm sfc}$ using the brentq method implemented in scipy \citep{scipy} (function root\_scalar), and use the value at which these fluxes coincide for $F_{p}$.

\subsubsection{Convective-radiative atmosphere}
\label{subsubsec:atmosphere}
We compute the radiative heat flux through an atmosphere composed of H$_2$, H$_2$O, CO, and CO$_2$. Similar to previous studies \citep{Marcq2017,Katyal2019}, the atmosphere consists (from the surface to the top of the atmosphere) of a dry troposphere, a moist troposphere (where water condenses), and a stratosphere. In the dry troposphere, the temperature profile is given by the dry adiabat:
\begin{equation}
    \frac{\partial\ln{T}}{\partial\ln{p}}=\frac{R}{\bar{c}_p},
    \label{eq:dry_adiabat}
\end{equation}
where $p$ is the gas (hydrostatic) pressure and $\bar{c}_p$ is the average molar heat capacity of the gas.
When the saturation pressure of water at the ambient temperature reaches the partial pressure of water, condensation occurs and the temperature gradient is given by the moist pseudo-adiabat (assuming complete rain-out of the condensate \citep{Pierrehumbert2010}):
\begin{equation}
    \frac{d\ln{T}}{d\ln{p_a}} = \frac{R}{c_{p,a}}\frac{1+\frac{L_{{\rm H}_2{\rm O}}}{RT}\frac{p_{{\rm H}_2{\rm O}}}{p_a}}{1+\left(\frac{c_{p,{\rm H}_2{\rm O}}}{c_{p,a}}+\left(\frac{L_{{\rm H}_2{\rm O}}}{RT}-1\right)\frac{L_{{\rm H}_2{\rm O}}}{c_{p,a}T}\right)\frac{p_{{\rm H}_2{\rm O}}}{p_a}},
    \label{eq:moist_adiabat}
\end{equation}
where $p_a$ is the sum of the partial pressures of all gases other than water (dry background air), $c_{p,a}$ the molar heat capacity of the dry background air, $L_{{\rm H}_2{\rm O}}$ and $c_{p,{\rm H}_2{\rm O}}$ the latent heat of vaporization and molar heat capacity of water, respectively. Equation \ref{eq:moist_adiabat} gives the slope of the (pseudo) adiabat (lapse rate) as a derivative in $p_a$ rather than total pressure. The system is closed by the Clausius-Clapeyron relation, which gives the partial pressure of water as a function of temperature (assuming that water is always at saturation in the moist troposphere, see Appendix \ref{apx:p_sat} for the expression of the saturation pressure). Equations \ref{eq:dry_adiabat} and \ref{eq:moist_adiabat} are solved using a Runge-Kutta 4$^{\rm th}$ order total- and dry air pressure-stepping, respectively, with $T_{\rm sfc}$ as boundary condition.

Finally, following previous studies \citep{Marcq2017,Katyal2019}, we consider an isothermal stratosphere at 200 K extending from the pressure level where the tropospheric temperature reaches 200 K up to the top of the atmosphere (defined at 1 Pa). While we assume that water is the only condensing species, we verify \textit{a posteriori} this assumption by comparing the partial pressures of all other gases present with their respective saturation pressures at ambient temperature, and found that in all cases water is indeed the only saturated gas.

The resulting temperature profile and mass mixing ratios are then used to compute the bolometric flux (outgoing longwave radiation, OLR) using the radiative transfer code petitRADTRANS \citep{Molliere2019}. The heat flux is computed as the wavelength-integrated flux (between 0.1 and 251 $\mu$m) using the correlated-$k$ method \citep{LacisOinas1991} with a coarse wavelength binning (10 bins). The coarse binning allows for rapid calculation as we do not need the wavelength-resolved spectrum. We also account for the collisional opacity of H$_2$, as it has been shown to be significant for high partial pressures of H$_2$ in MO atmospheres \citep{Lichtenberg2021}.

\begin{table}[h!]
    \centering
    \begin{tabular}{l l l l}
        {\bf Symbol} & {\bf Quantity} & {\bf Value/Expression} & {\bf Units} \\ \hline
        $E_{\rm th}$ & thermal energy of the planet & evolving & J \\
        $dE_{\rm th}/dT_{\rm pot}$ & secular cooling & $8.34\times10^{27}$ & J/K \\
        $M_{\rm sol}$ & mass of solid in the mantle & evolving & kg \\
        $L$ & specific latent heat of crystallization & 500 & J/kg \\
        $H_{\rm int}$ & radioactive heating & $1.31\times10^{14}$ & W \\
        $R_p$ & planetary radius & $6371\times10^3$ &  m\\
        $F_p$ & heat flux through planetary surface & evolving & W/m$^2$ \\
        $F_{\rm conv}$ & MO convective heat flux & evolving & W/m$^2$ \\
        $F_{\rm rad}$ & atmospheric radiative heat flux & evolving & W/m$^2$ \\
        $d$ & MO depth & evolving & m \\
        $k$ & thermal conductivity & $\kappa/\rho/c_p$ & J/m/K \\
        Ra & MO Rayleigh number & Eq. \ref{eq:Rayleigh_number} & - \\
        $\kappa$ & thermal diffusivity & $10^{-1}$ & m$^2$/s \\
        $\eta$ & viscosity & Eq. \ref{eq:VFT} & Pa s \\
        $T_{\rm sfc}$ & MO surface temperature & evolving & K \\
        $p$ & hydrostatic pressure & variable & Pa\\
        $\bar{c}_p$ & average molar heat capacity in the dry troposphere & variable & J/K/mol \\
        Pr & Prandtl number & $\eta/(\rho\kappa)$ & - \\
        $\lambda$ & convective flow aspect ratio & 1 & - \\
        $c_{p,a}$ & molar heat capacity of the dry gas in the dry troposphere & variable & J/K/mol \\
        $c_{p,{\rm H}_2{\rm O}}$ & molar heat capacity of water & \citet{JANAF} & J/K/mol \\
        $p_a$ & partial pressure of the dry gas in the moist troposphere & variable & Pa \\
        $L_{{\rm H}_2{\rm O}}$ & molar latent heat of condensation of water & 44.874 & kJ/mol \\
        $R$ & gas constant & 8.314 & J/K/mol \\ 
        \hline
    \end{tabular}
    \caption{Notation in use in Section \ref{subsec:thermal_model}.}
    \label{tab:param3}
\end{table}

\section{Results}
\label{sec:results}
We calculate the evolution of the $f_{{\rm O}_2}$ in the MO, the corresponding redox speciation of the outgassed C-O-H atmosphere, and the thermal evolution of the MO cooling through this greenhouse atmosphere. We investigate the two crystallization scenarios presented in Sections \ref{subsubsec:equilibrium_crystallization} and \ref{subsubsec:fractional_crystallization} and the two initial conditions on $f_{{\rm O}_2}$ discussed in Section \ref{subsubsec:fO2_in_MO}. The initial volatile budget of the terrestrial MO is poorly constrained. \citet{Marty2012} and \citet{Hirschmann2018} provide estimates for the bulk silicate Earth (BSE) volatile content (11.6 and 2 EO-equivalent of H, respectively, and both find C/H$\sim1.7$), but it is still debated whether these volatiles were delivered during the late veneer \citep{Albarede2009} or were already present in the building blocks of the Earth \citep{Piani2019,Grewal2021}. In the former case, estimates of the BSE volatile content do not reflect the content in the terrestrial MO, while in the latter case they provide a lower bound. \citet{CatlingKasting2017} (p. 157) suggests that, accouting for the time-evolution of the sun's XUV flux, if energy-limited escape had been operating throughout Earth's life (which is not realistic but provides and upper bound of the escape flux), it would have depleted about 10 Earth ocean (EO)-worth of H.
As this provides a generous overestimate of the amount of H lost, 
here we vary the H  budget of the BSE budget between 1 and 9 EO of water. In order to cover a large parameter space, we vary the mass C/H ratio between 0.5 and 2. Notice that the volatile budgets are indicated as the BSE contents, i.e. contents in the combined (initial) MO and unmolten lower mantle, assuming a homogeneous distribution (as can be expected if volatiles where delivered during Earth's accretion). Thus, since the MO represents initially $\sim58$ wt\% of the mantle, the actual MO volatile content is in the same proportion (e.g. 0.58 Earth water ocean-worth of H content for the minimum H end-member). The volatile content in the lower mantle is not relevant to the present study.

\subsection{Evolution of $f_{{\rm O}_2}$ during magma ocean crystallization}
\label{subsubsec:results:fO2}
\begin{figure}[h!]
    \centering
    \includegraphics[width=\textwidth]{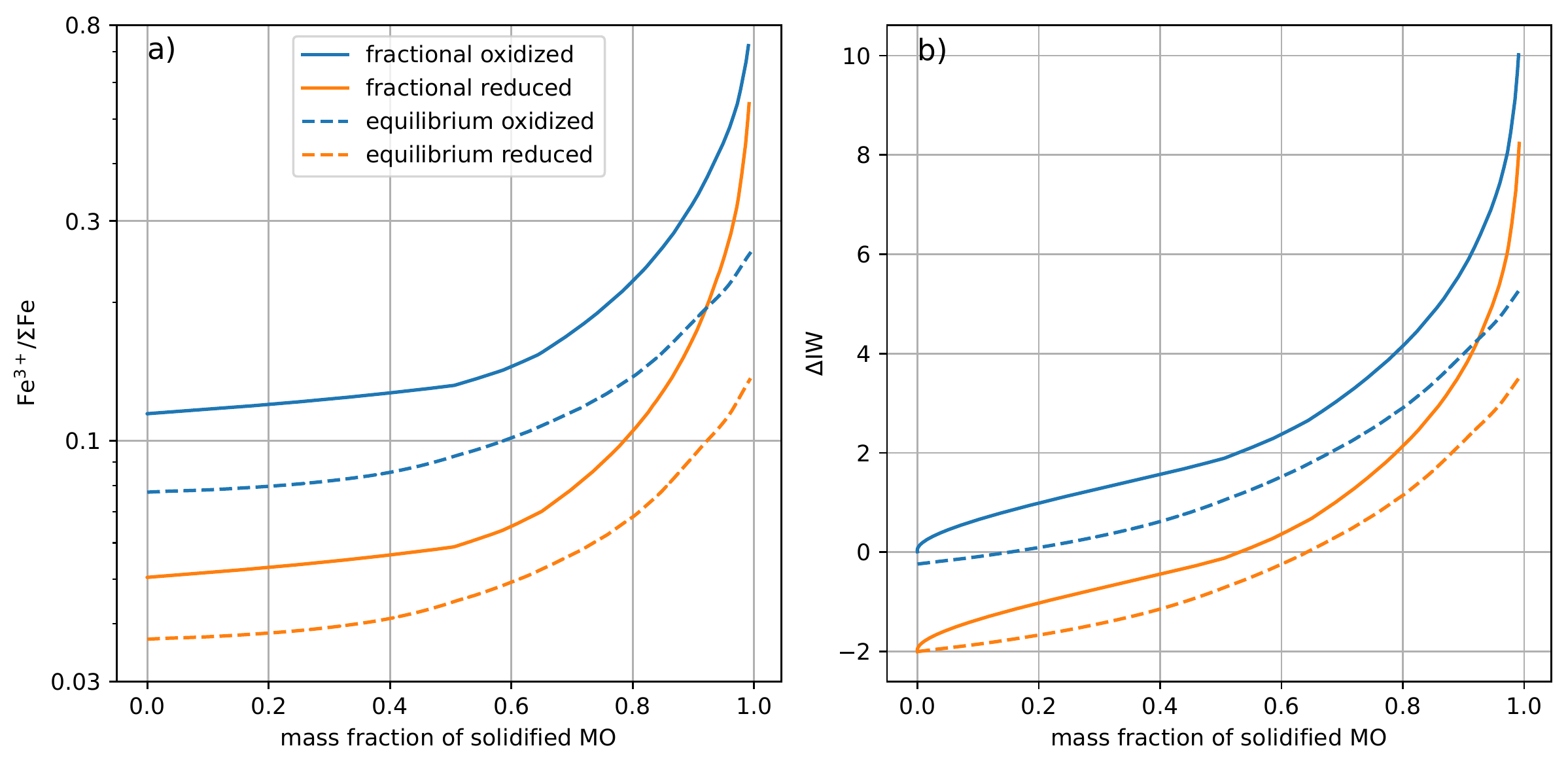}
    \caption{Evolution of ${\rm Fe}^{3+}/\Sigma{\rm Fe}$ (a) and the surface $f_{{\rm O}_2}$ (b) throughout MO crystallization. On both panels, solid curves correspond to fractional crystallization and dashed curves to equilibrium crystallization, while orange curves correspond to an initial $\Delta{\rm IW}=-2$ at the surface (reduced cases) and blue curves to an initial $\Delta{\rm IW}=-2$ at the bottom of the MO (resulting in a higher surface $f_{{\rm O}_2}$, hence oxidized cases). Notice that the fraction of solidified MO ($x$-axis) is different from the crystals fraction in the case of equilibrium crystallization because the former  includes the melt trapped in the solid cumulates while the latter does not.}
    \label{fig:Fe3+_fO2}
\end{figure}

We calculate the evolution of ${\rm Fe}^{3+}/\Sigma{\rm Fe}$ in the liquid throughout MO crystallization for both crystallization scenarios, and both initial $f_{{\rm O}_2}$ conditions (Figure \ref{fig:Fe3+_fO2}a). The different initial conditions on $f_{{\rm O}_2}$ translate to different initial conditions for ${\rm Fe}^{3+}/\Sigma{\rm Fe}$ (see sections \ref{subsubsec:fO2_in_MO} and \ref{apx:initial_redox_state}). The difference in initial ${\rm Fe}^{3+}/\Sigma{\rm Fe}$ between the two crystallization scenarios is due to the difference in initial potential temperature, and thus the $P$-$T$ conditions at which $\Delta{\rm IW}=-2$ (see Figure \ref{fig:initial_redox_profiles}b).
The higher incompatibility of ${\rm Fe}^{3+}$ relative to ${\rm Fe}^{2+}$ leads to an increase of ${\rm Fe}^{3+}/\Sigma{\rm Fe}$ in the liquid as the MO crystallizes for all cases. Fractionation is stronger for fractional crystallization because no melt is retained in the solid cumulates, while they retain a fraction $\phi_{\rm RCMF}$ in equilibrium crystallization. Thus, a significantly higher ${\rm Fe}^{3+}/\Sigma{\rm Fe}$ is reached at the end of fractional crystallization cases.

Using liquid's ${\rm Fe}^{3+}/\Sigma{\rm Fe}$, we compute the surface $f_{{\rm O}_2}$ (evaluated at $T_{\rm pot}$) during MO crystallization (Figure \ref{fig:Fe3+_fO2}b). \citet{Sossi2020} posits that the temperature dependence of $f_{{\rm O}_2}$ as set by reactions \ref{eq:iron-wuestite_equilibrium} and \ref{eq:ferrous-ferric-iron_equilibrium} are similar, which would dismiss the influence of temperature on $\Delta{\rm IW}$. However, \citet{Hirschmann2022} argues that taking into consideration the difference in heat capacity between FeO and FeO$_{1.5}$, Reaction \ref{eq:ferrous-ferric-iron_equilibrium} has an increased temperature dependence past 2000 K. As a matter of fact, under a pressure of 1 bar  and for a fixed ${\rm Fe}^{3+}/\Sigma{\rm Fe}$, $\Delta{\rm IW}$ as calculated from \citet{Hirschmann2022}'s parametrization increases by $\sim$3 log-units upon decreasing the temperature from 3000 K to 1500 K (roughly the range addressed in the present study). Thus the trend observed in Figure \ref{fig:Fe3+_fO2}b is due to both increase in Fe$^{3+}/\Sigma$Fe as well as the decrease in potential temperature as the MO crystallizes.

The effects of the temperature and of ${\rm Fe}^{3+}/\Sigma{\rm Fe}$ on $f_{{\rm O}_2}$ act in the same direction and yield a final increase of $\Delta{\rm IW}$ of several log-units in all cases. The equilibrium crystallization cases (dashed curves in Figure \ref{fig:Fe3+_fO2}) exhibit the lowest increase in ${\rm Fe}^{3+}/\Sigma{\rm Fe}$, and most of their increase in $\Delta{\rm IW}$ is accounted for by the decrease in $T_{\rm pot}$. Conversely, in the fractional crystallization scenario (solid curves in Figure \ref{fig:Fe3+_fO2}), ${\rm Fe}^{3+}/\Sigma{\rm Fe}$ reaches high values (FeO$_{1.5}$ becoming the major iron oxide) and induces a highly oxidized final state.

\subsection{Atmosphere speciation and magma ocean volatile content}
\label{subsec:atmosphere_speciation}

The oxidation state of the MO obtained in the previous section is used to calculate the composition of the outgassed atmosphere. Three main effects govern the evolution of atmospheric pressure and composition during MO crystallization.

First, as both H and C are incompatible elements, they are partitioned into silicate melts during crystallization. Thus, MO solidification increases their concentrations in the residual melt. This increase is accommodated by outgassing until the partial pressures of the volatile-bearing species in the atmosphere and their concentration in the silicate melt reach an equilibrium given by solubility laws (Equation \ref{eq:Henry_law}, assumed to be an instantaneous process).
Second, when an element exists in different species in the gaseous phase (e.g. oxidized or reduced), the evolution of its outgassing is further complicated by the different solubilities of the gaseous species. For instance, in the H-bearing redox system, ${\rm H}_2{\rm O}$ has a significantly higher solubility in silicate melts than ${\rm H}_2$, allowing for more H to dissolve when the $f_{{\rm O}_2}$ in the MO increases, all other things being equal. 
Third, when several species are present in the atmosphere, variations in the partial pressure of one of them influence the others: the average molecular weight in the atmosphere is altered, which modifies the relationship between the mass of one species in the atmosphere and its partial pressure (as per equation \ref{eq:mass_cons}). If the average molecular mass of the atmosphere decreases (for instance if H$_2$O is outgassed in a CO$_2$-dominated atmosphere), the partial pressures of other species will decrease to accomodate the same mass in the atmosphere (e.g. \citet{Bower2019}). Actually, this effect is associated with a mass redistribution between the atmosphere and the MO, since a modification of the partial pressure induces a modification of the solubility and hence of the dissolved mass. This leads to the surprising configuration where a reduction of the partial pressure of a gas is correlated with its net outgassing, because its solubility is accordingly decreased.

\begin{figure}[h!]
    \centering
    \includegraphics[width=\textwidth]{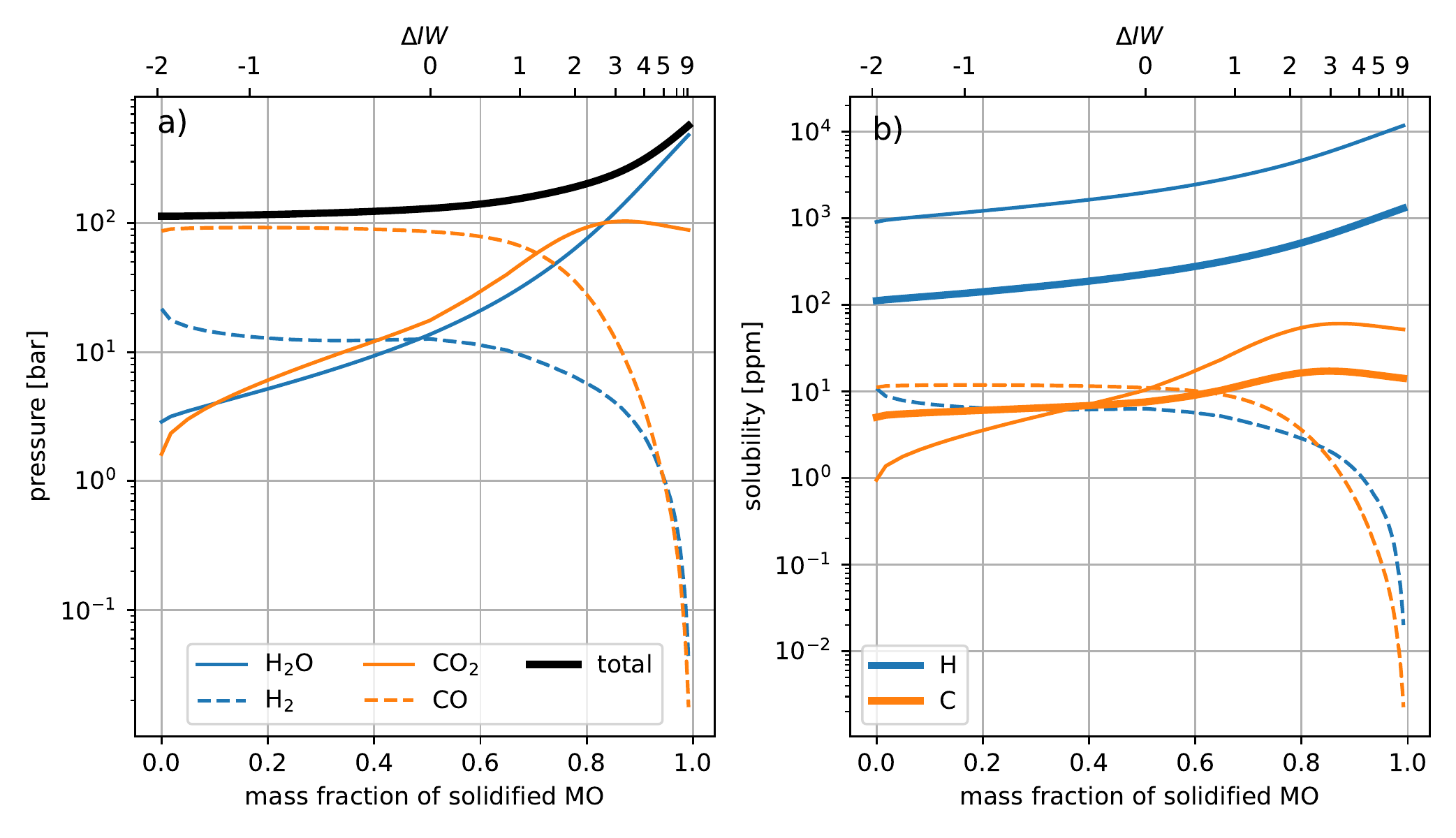}
    \caption{Evolution: (a) of the partial and total pressures and (b) the solubilities throughout MO solidification. In panel b, thick curves represent the total solubility of an element while the thin curves represent the species-specific solubilities (corresponding to panel a's legend). The surface $\Delta{\rm IW}$ at the corresponding fraction of solidified MO is reported on the top $x$-axis.}
    \label{fig:pressures_solubilities}
\end{figure}

How these three effects operate is controlled by specific features of the H and C systems, in particular the difference in solubility between oxidized and reduced species and the temperature dependence of the equilibrium constant of their respective redox reactions. For a BSE content of 3 EO-worth of H and a C/H ratio of 1, undergoing fractional crystallization, with an initial $f_{{\rm O}_2}$ buffered at IW-2 at the surface (reduced case), Figure \ref{fig:pressures_solubilities}a represents the evolution of the partial pressures of the different volatile-bearing species in the atmosphere, and Figure \ref{fig:pressures_solubilities}b the solubilities of H and C. We chose this case to illustrate the evolution of the volatile speciation because its $f_{{\rm O}_2}$ crosses the stability domains of most species, contrary to the more oxidized cases.

In both C and H redox systems, the oxidized species has a higher solubility than the reduced one. Thus the increase in $f_{{\rm O}_2}$ with MO crystallization reported in the previous section competes with the partitioning of volatiles in the remaining melt: the former reduces outgassing (first effect mentioned above) while the latter promotes it (second effect mentioned above). However, in the case of C, both species have a fairly low solubility, and most C is readily outgassed since the beginning of MO solidification. The effect of redox speciation on C solubility - as accounted for by our model - is thus marginal. The possibility of graphite saturation (that would reduce C solubility and sequester a part of the C budget) is discussed in Section \ref{subsec:graphite_saturation}. The solubility difference between oxidized and reduced species is much larger in the case of the H system, where water has a solubility in silicate melt higher than molecular hydrogen by several orders of magnitude \citep{Hirschmann2016}. Even at the beginning of the crystallization, when the MO is the most reduced and H$_2$ is ten times more abundant than water in the atmosphere, it represents only $\sim$ 10\% of the total dissolved H (Figure \ref{fig:pressures_solubilities}b). Thus the high solubility of water compensates for the unfavourable redox conditions and the effective solubility of H is set by water at all times.

Contrarily to other species whose solubility follow Henry laws, H$_2$O's partial pressure increases as its solubility to the power $1/\beta_{{\rm H}_2{\rm O}}>1$. Thus, partitioning of water in the remaining MO leads to a  sharper increase in $p_{{\rm H}_2{\rm O}}$ compared to $p_{{\rm CO}_2}$ (the other species favoured by redox evolution) in the late part of the crystallization. In particular, following the third effect mentioned above, the delayed outgassing of water decreases the partial pressure of CO$_2$ in the last 20\% (see flattening and even final decrease of the solid orange curve in Figure \ref{fig:pressures_solubilities}a).

\begin{figure}[h!]
    \centering
    \includegraphics[width=\textwidth]{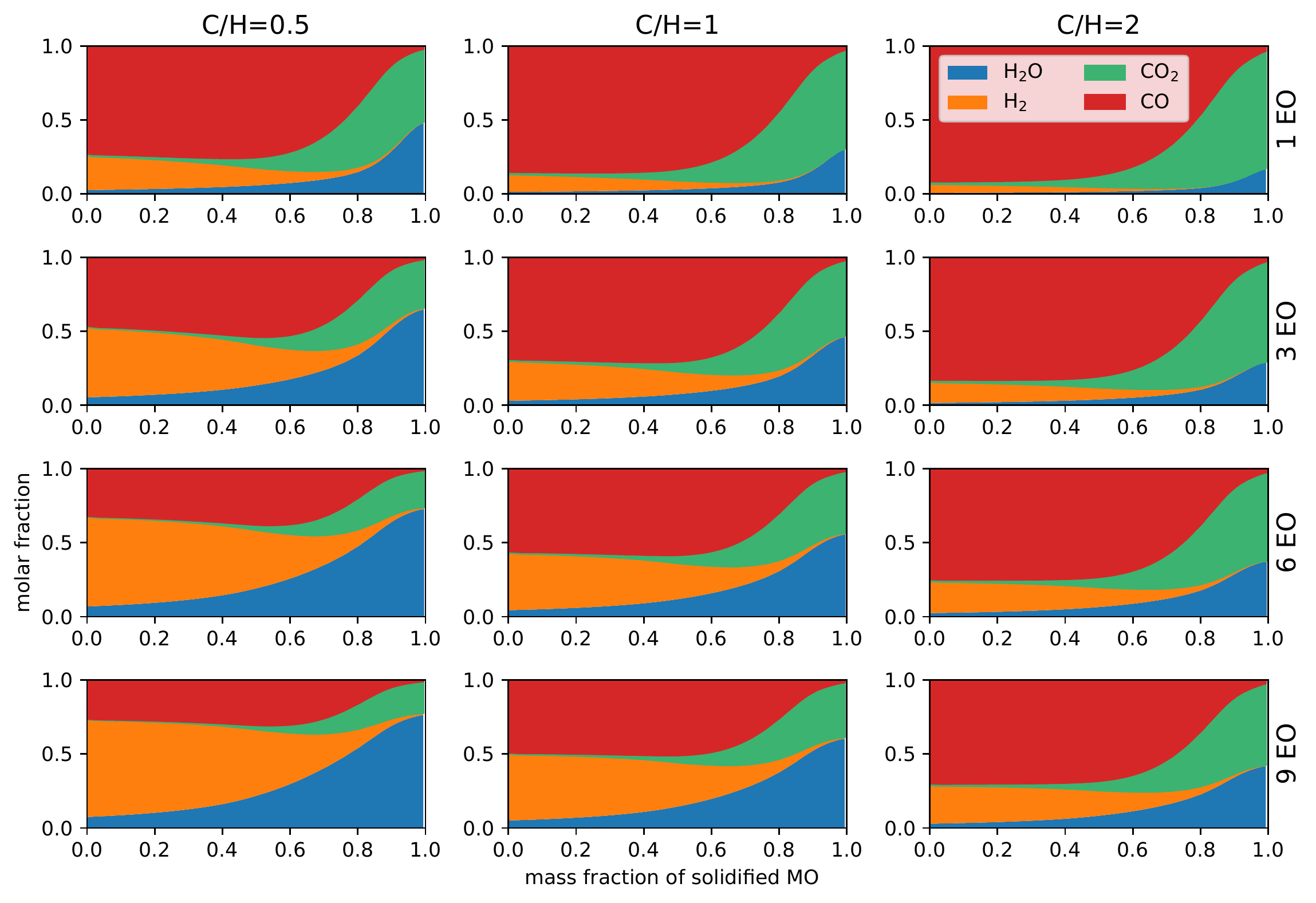}
    \caption{Evolution of the molar mixing ratios for BSE H contents between 1 and 9 EO-worth of H and C/H ratios from 0.5 to 2 for a MO undergoing equilibrium crystallization (initial $\Delta{\rm IW}=-2$ at the surface).}
    \label{fig:bulk_H_and_C_to_H_frac}
\end{figure}

The initial BSE abundances of C and H influence the evolution of the MO outgassing. Figure \ref{fig:bulk_H_and_C_to_H_frac} represents the evolution of the molar mixing ratios of the atmospheric species throughout the MO crystallization over all the combinations of BSE H budgets and C/H ratio, for a ``reduced'' MO undergoing fractional crystallization. Increasing the C and H mass budgets by the same amount (i.e. keeping C/H constant) induces a significant increase in H$_2$'s molar mixing ratio. This is due to the small molecular mass of this species: the same increase in mass in the atmosphere corresponds to a larger increase in partial pressure for light species than for heavy ones. Perfect fractional crystallization being assumed in this case, all volatiles are eventually outgassed. Since all cases represented follow the same $f_{{\rm O}_2}$ evolution where the final redox state is oxidized enough for reduced species to be negligible, they all converge toward similar final atmospheric compositions, where the relative molar fractions only depend on C/H.

As already pointed out in the case of iron oxides partitioning, enrichment in incompatible elements in the MO is much lower in the case of equilibrium crystallization compared to fractional crystallization (Figure \ref{fig:Fe3+_fO2}a). This effect is even stronger for volatiles, which we assume to be perfectly incompatible. It follows that volatiles are significantly less outgassed in the former case, and a large amount is retained in melts trapped in the solid cumulates. This applies particularly to H, due to the high solubility of ${\rm H}_2{\rm O}$. Furthermore, the lower final $f_{{\rm O}_2}$ reached in the case of equilibrium crystallization induces a more reduced final atmosphere (although still dominated by H$_2$O and CO$_2$ for all cases).

\begin{figure}[h!]
    \centering
    \includegraphics[width=\textwidth]{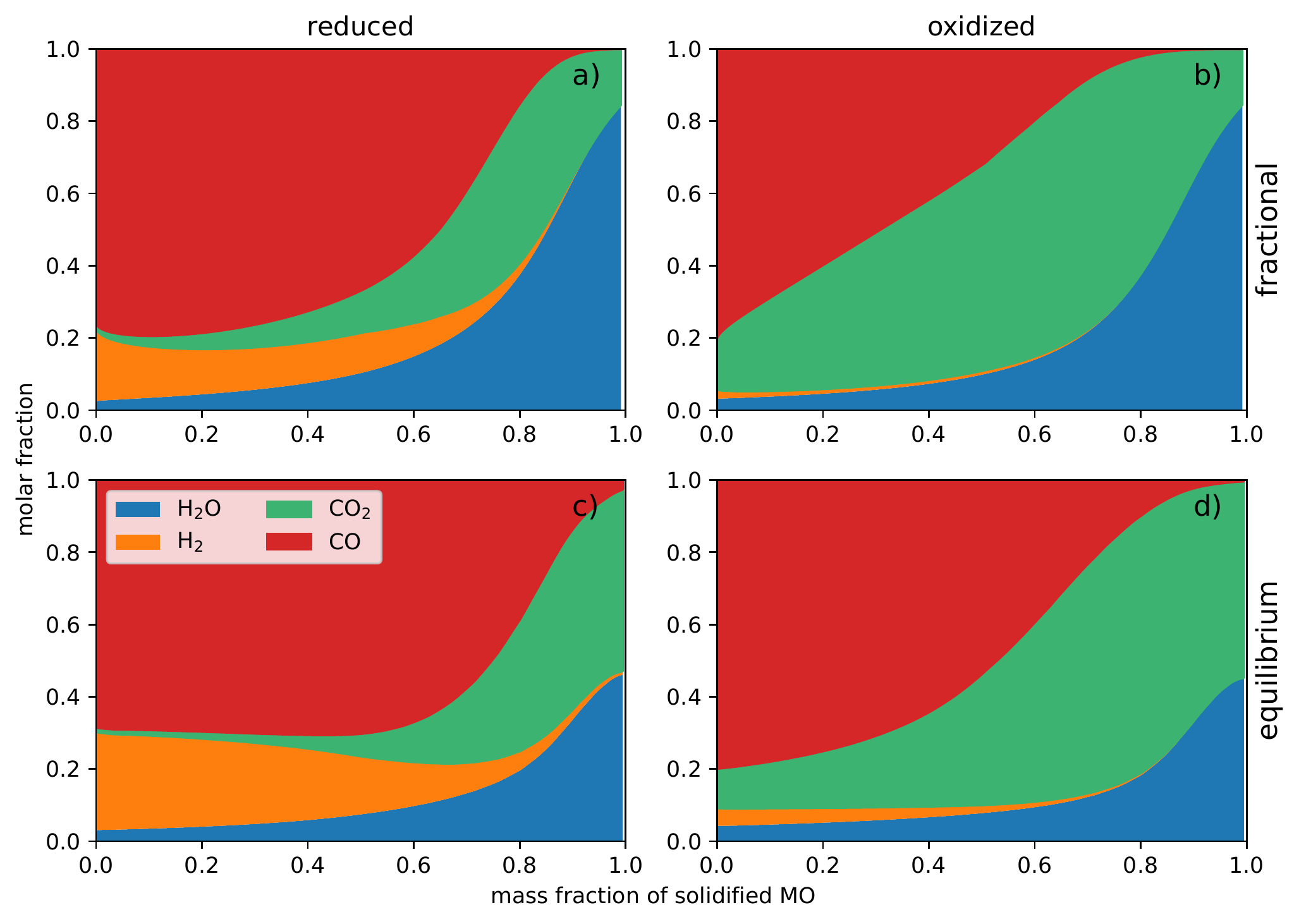}
    \caption{Molar fraction evolutions (colored areas, read off left $y$-axis) and average molecular mass in the atmosphere (black curve, read off right $y$-axis) for a case with 3 EO-worth of H in the BSE and C/H=1, for fractional (upper row) and equilibrium (lower row) crystallization and initially reduced (left column) and oxidized (right column) MO.}
    \label{fig:frac_equi_red_ox}
\end{figure}

Figure \ref{fig:frac_equi_red_ox} represents the evolution of the atmospheric composition for both crystallization scenarios and both initial redox states, for a case with the same volatile inventories as that presented in Figure \ref{fig:pressures_solubilities}.
The late outgassing of water induces the final decrease in the average molecular mass in the atmosphere (black curves), observed in all cases.
However, the final mixing fraction of water is much smaller in the equilibrium crystallization scenario, because a large quantity remains sequestered in the trapped melts. This translates into a milder final decrease of the molecular mass in the atmosphere, as well as a dicothomy in the net change in atmospheric molecular mass over MO solidification: negative in the case of fractional crystallization (because H$_2$O, the dominant gas in the final atmosphere, is lighter than CO, the dominant gas in the initial atmosphere), and positive in equilibrium crystallization cases, whose final atmospheres are dominated by heavy CO$_2$.

An oxidized initial state leads to a drastic reduction of H$_2$ and an increase of CO$_2$ (H$_2$O staying largely dissolved in the MO), and thus an increased solubility of H compared to C. Because C-bearing gases are heavier than H-bearing ones, this translates into an increased average molecular mas of the atmosphere (compare black curves in each column).

\subsection{Timescale of Magma Ocean Crystallization}
\label{subsec:thermal_evolution}

Previous studies found that MO outgassing of greenhouse species (in particular ${\rm H}_2{\rm O}$ and ${\rm CO}_2$) increases the opacity in the atmosphere, thereby decreasing the cooling flux of the MO and prolonging its lifetime \citep{AbeMatsui1985,AbeMatsui1988,ElkinsTanton2008,Lebrun2013,Salvador2017,Nikolaou2019,Bower2019}. As shown in the previous section, redox speciation in the atmosphere influences volatiles solubilities, and thus atmosphere outgassing. Furthermore, the different species of the redox couples have different radiative as well as thermodynamic properties, affecting the infrared opacity and lapse rate of the atmosphere. Therefore, redox speciation in the atmosphere further influences the time-evolution of MO crystallization. 

\begin{figure}[h!]
    \centering
    \includegraphics[width=\textwidth]{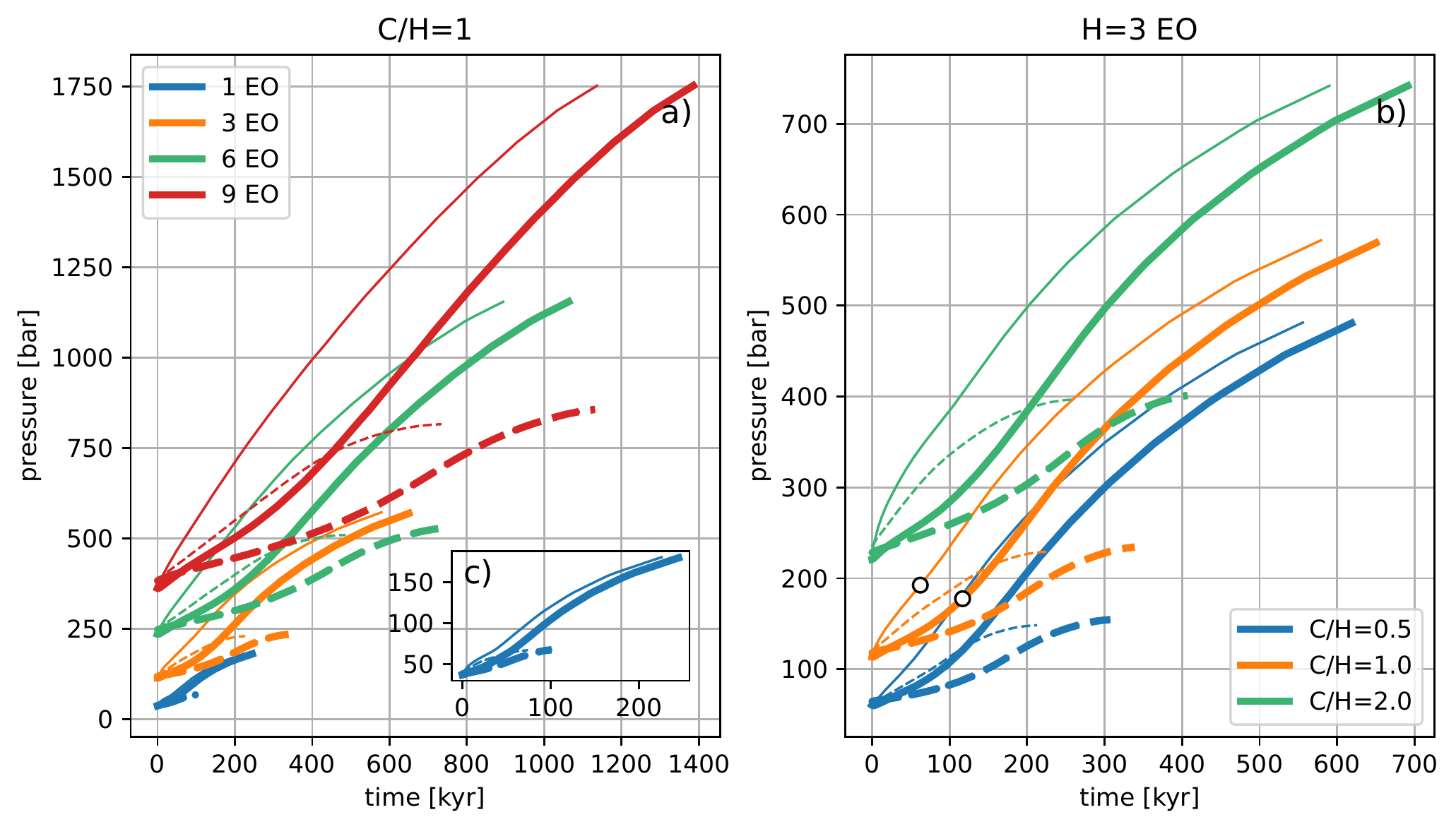}
    \caption{Time-series of the total pressure for all H BSE budgets keeping C/H constant at 1 (a), and for all C/H values keeping the BSE H budget constant to 3 EO (b). Panel c is a blow-up of the cases having the a BSE H content of 1 EO. Thick and thin lines represent reduced and oxidized cases, respectively, while solid and dashed lines correspond to fractional and equilibrium crystallization cases, respectively. Orange curves coincide on both panels (notice that time axis differ between the panels. White circles correspond to the 75 wt\% MO solidification, reported in the middle panel b of Figure \ref{fig:red_vs_ox_OLR} (panels a and c corresponding to the origin and the end of the corresponding lines, respectively).}
    \label{fig:thermal_evol}
\end{figure}

Figure \ref{fig:thermal_evol} represents the time evolution of the total atmospheric pressure for all investigated values of the C/H ratio, keeping a BSE H budget of 3 EO (panel a), as well as for all the BSE H budgets envisaged, keeping C/H=1 (panel b). Among the oxidized species, ${\rm H}_2{\rm O}$ is a stronger infrared absorber than ${\rm CO}_2$. Among the reduced ones, H$_2$ and CO are only mild greenhouse gases, but \citet{Lichtenberg2021} found that H$_2$ collision-induced absorption might be significant. It may thus be anticipated that H will be more influential than C in setting the MO lifetime, and hence varying the H budget will be crucial. This is verified in Figure \ref{fig:thermal_evol}a, where increasing the H budget from 1 to 9 Earth ocean masses of water prolongs the MO crystallization by $\sim1$ order of magnitude. We notice that accounting for the H$_2$-H$_2$ collision-induced absorption does not significantly prolong the MO lifetime, even in cases where the partial pressure of H$_2$ is the highest ($\sim 145$ bar). \citet{Lichtenberg2021} pointed out the important thermal blanketing effect of a thick H$_2$ atmosphere, but for a higher H$_2$ pressure ($\sim200$ bar). The influence of C/H (Figure \ref{fig:thermal_evol}b) also follows an intuitive trend, whereby increasing the C/H ratio while keeping the H inventory constant (i.e. increasing the C inventory) translates into a prolongation of the MO lifetime. This prolongation is only marginal compared to that induced by increasing the H budget.
Furthermore, as already observed, fractional crystallization of the MO allows for a much more extensive outgassing than equilibrium crystallization; thus in turn, for a slower cooling. Indeed, we consistently observe that, all other things being equal, the former (solid curves) yields significantly longer MO than the latter (dashed curves) in Figure \ref{fig:thermal_evol}. Over the whole parameter space covered, the shortest MO (initially reduced MO undergoing equilibrium crystallization, with a BSE H budget of 1 EO and C/H=0.5) solidifies in 67 kyr, and the longest MO (initially oxidized MO undergoing fractional crystallization, with a BSE H budget of 9 EO and C/H=2) solidifies in 1.530 Myr.
\begin{figure}[h!]
    \centering
    \includegraphics[width=\textwidth]{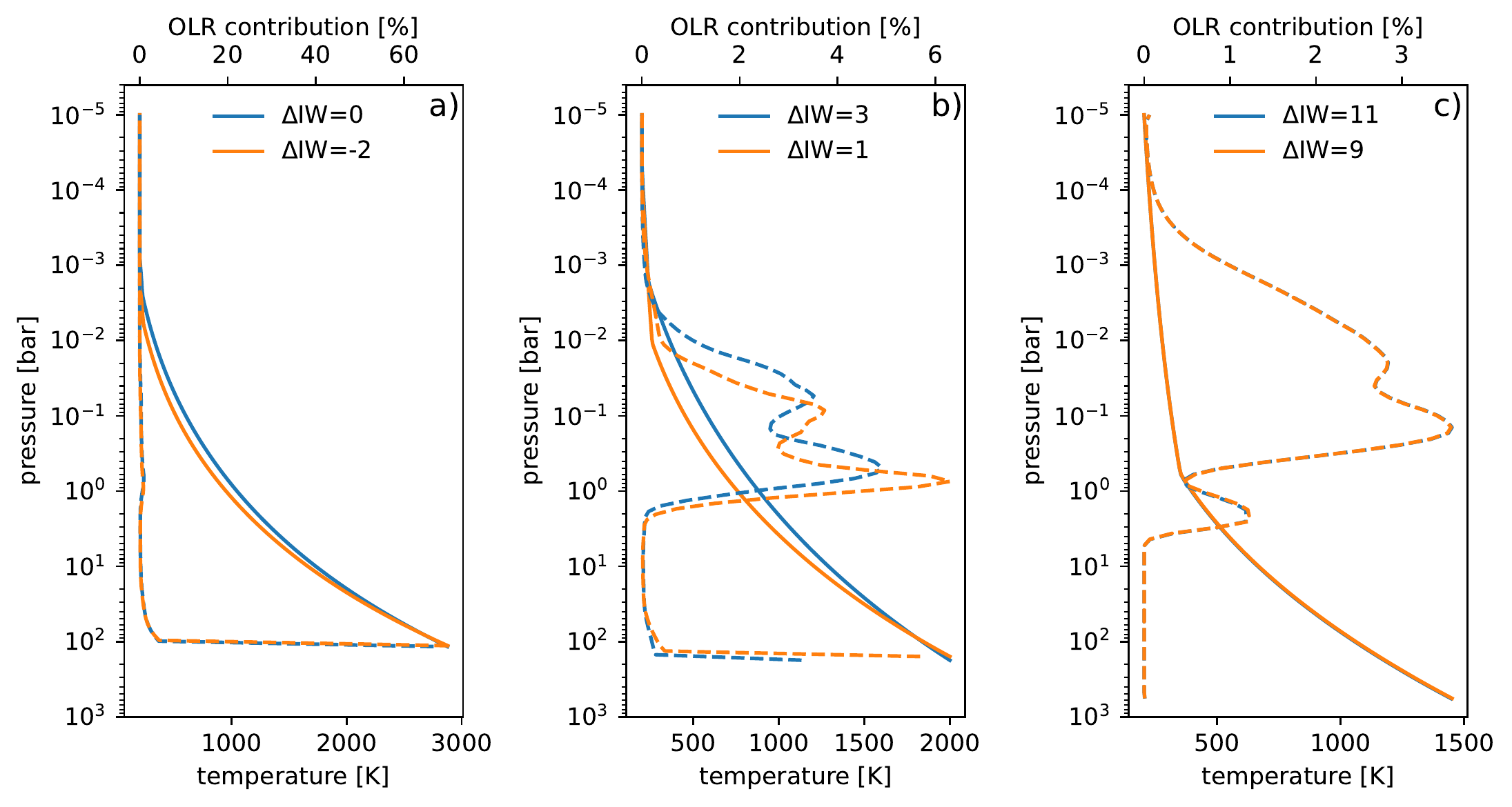}
    \caption{Profiles of temperature (solid lines, read off bottom $x$-axis) and OLR contributions (dashed lines, read off top $x$-axis) for atmospheres overlying an initially oxidized (blue lines) and an initially reduced (orange lines) MO, after 0 wt\% (panel a), 75 wt\% (panel b) and 98 wt\% (panel c) of the MO solidified. The $f_{{\rm O}_2}$ of the atmosphere at each corresponding stage is indicated in the legend. Notice that the $y$-axis are the same for all three panels, while their $x$-axis (both top and bottom) vary from one panel to the other.}
    \label{fig:red_vs_ox_OLR}
\end{figure}

One noticeable feature in Figure \ref{fig:thermal_evol} is the consistently longer lifetime of initially reduced MOs (thick lines) compared to initially oxidized ones (thin lines). This effect seems to contradict the stronger infrared absorption of oxidized gases compared to reduced ones. Indeed, \citet{Katyal2019} found that, for a fixed snapshot in the MO evolution, reduced cases exhibit larger OLR than oxidized ones, which should lead to their shorter lifetimes. The surface temperature for which they calculated the OLR correspond to the beginning (3000 K) and the end (1500 K) of MO crystallization (these temperatures are roughly similar to the ones in the present study). As described below, we do observe that the OLR of oxidized cases become larger than that of reduced ones for intermediate to high temperature, but are similar for such end-member temperatures. \citet{Bower2022} also calculated the OLR for both reduced and oxidized MO atmospheres, and found the former to crystallize faster than the later (in agreement with \citet{Katyal2020}'s results). However, the simple atmosphere model they used (from \citet{AbeMatsui1985}) cannot capture the influence of atmospheric chemistry on the temperature profile, which we show is key to the effect responsible for the stronger radiation of oxidized atmospheres.

The radiative heat flux through the atmosphere is the result of species-specific (wavelength-dependent) absorption features and temperature-controlled emission of all pressure levels. For a fractional crystallization case (BSE H budget of 3 EO, C/H=1), Figure \ref{fig:red_vs_ox_OLR} represents the temperature and OLR contribution profiles in the atmosphere at three successive snapshots of MO crystallization: 1) beginning of crystallization, 2) after 75 wt\% of the MO has crystallized (corresponding to a surface temperature of 2000 K, represented by white circles in Figure \ref{fig:thermal_evol}b), and 3) end of MO crystallization. The OLR contribution is the normalized, wavelength-integrated contribution of each pressure level to the OLR. Generally, at low (sub-solidus) temperature, emission from the deepest layers of thick ($>$10 bar) atmospheres is fully absorbed, and only shallower layers' emission effectively contributes to the OLR. However here, the extremely hot surface of the MO does contribute significantly to the initial OLR, despite the relatively thick atmosphere. Figure \ref{fig:red_vs_ox_OLR}a shows that the surface of the MO accounts for 75\% of the OLR at the beginning of MO crystallization.

As crystallization proceeds, the surface contribution diminishes and the bulk OLR contribution  is offset towards higher levels (Figure \ref{fig:red_vs_ox_OLR}b). While the height of the radiating levels is controlled by the absorption of overlying layers (and thus by the composition of the atmosphere), the intensity of their radiation is governed by their temperature, which in turn is set by the lapse rate in the troposphere (the stratospheric contribution being always negligible). Among the gaseous species considered here, CO$_2$ and H$_2$O, which dominate the oxidized phase of MO crystallization, have the highest molar heat capacities, while CO and H$_2$, more important in the reduced phase, have molar heat capacity values about half that of their oxidized counterparts. Hence, reduced atmospheres exhibit a stronger lapse rate and a colder upper atmosphere (Figure \ref{fig:red_vs_ox_OLR}b). Although the bulk of the OLR-contributing pressure levels lies at slightly lower $p$ in the oxidized case, their temperature is higher than in the reduced case due to the difference in the lapse rate, resulting in lower OLR of the latter case.

Eventually, as the MO crystallizes, both cases evolve toward a strongly oxidized atmosphere, and they become almost indistinguishable. At the same time, the surface temperature further cools down and the atmosphere becomes thicker, so the contribution from the surface vanishes, and the main contributing layers spread over a larger pressure range, mostly in the moist troposphere (Figure \ref{fig:red_vs_ox_OLR}c). The stronger OLR exhibited by oxidized atmospheres is consistent throughout the parameter space explored, resulting in a general shorter lifetime of initially oxidized MOs.

\section{Discussion}
\label{sec:discussion}

\subsection{Volatiles retention in the MO cumulates}
\label{subsec:volatile_mantle}
In the case of fractional crystallization, volatiles are virtually entirely outgassed (since we neglect partitionning of volatiles in solids). In the equilibrium crystallization cases, however, the large amount of interstitial melt trapped within the solid mantle allows for sequestration of a significant mass of volatiles. We expect chemical speciation to play a role in the retention of volatiles in the mantle, by indirectly setting the effective elemental solubilities. H in particular can be retained due to the high solubility of water. Figure \ref{fig:pressures_solubilities} showed that even in the reducing case, H solubility is set high by water. As a result, a large quantity of H is retained in the interstitial melt trapped in the mantle. In contrast, both CO and CO$_2$ have a low solubility, and C is mostly outgassed since the beginning of the MO crystallization. However, CO$_2$ has a solubility approximately twice that of CO, so the redox speciation must affect whatever little C is retained in the mantle.

\begin{figure}
    \centering
    \includegraphics[width=\textwidth]{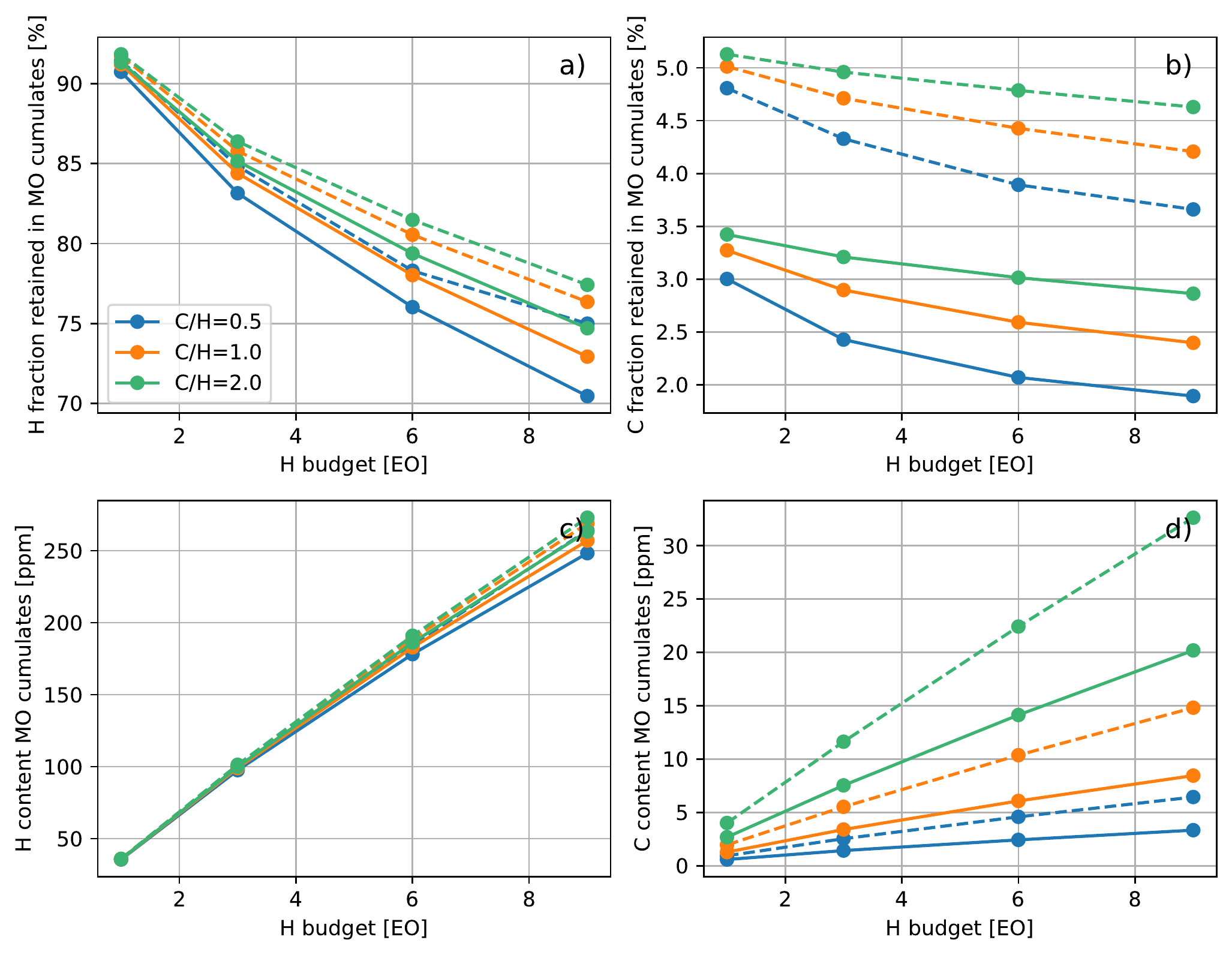}
    \caption{Fraction of volatiles initially in the MO+atmosphere retained in the final cumulates (H (a) and C (b)), and their corresponding contents in the solid cumulates (H (c) and C (d)), for all equilibrium crystallization cases. We note again that in our model the volatile retention in solid cumulates is in the form of trapped melts and not via incorporation in the crystal lattice.     Colors distinguish C/H while linestyle distinguish initial redox state (solid for reduced, dashed for oxidized).}
    \label{fig:volatiles_mantle}
\end{figure}

Figures \ref{fig:volatiles_mantle}a and b present the fraction of the initial MO volatiles mass (i.e. 58\% of the incidated BSE budget) that is retained in the solid cumulates of the MO over the whole parameter space (for equilibrium crystallization). We observe that for all cases, more than 70\% (and up to $\sim93\%$) of the H, and from $\sim1.8\%$ to $\sim5.2\%$ of C, are retained in the cumulates. Increasing the bulk quantity of H has an opposite effect to increasing the C budget on the retention of both elements in the MO cumulates. Increasing the bulk H budget decreases the final relative amount of both H and C retained. This is due to the corresponding increase in H$_2$ in the atmosphere reported in Section \ref{subsec:atmosphere_speciation} (see Figure \ref{fig:bulk_H_and_C_to_H_frac}), which promotes the outgassing of other species by lowering the average molecular mass of the atmosphere. On the contrary, increasing the bulk inventory of C leads to an increased volatiles retention in the MO cumulates because both CO and CO$_2$ are heavier than H$_2$O (and H$_2$), so increasing the bulk C (which is largely partitionned in the atmosphere) increases the average molecular mass of the atmosphere, deferring outgassing. Figures \ref{fig:volatiles_mantle}c and d present the final H and C contents of the MO cumulates. The small decrease in the fraction of retained volatiles with increasing BSE volatiles budget is largely offset by the absolute increase in BSE volatile budget. The H content in the cumulates increases from $\sim32$ ppm to $\sim270$ ppm as the BSE H budget increases from 1 to 9 EO. Similarly, the retained fraction of C increases from less than 1 ppm up to $\sim33$ ppm.

The oxidation state also plays an important role for C. As already pointed out, water sets the effective solubility of H, even in the most reduced phase of MO crystallization. For C, however, CO$_2$ solubility is about twice that of CO, so oxidized conditions promote C retention. Furthermore, the transition between a CO-dominated and a CO$_2$-dominated atmosphere occurs after approximately 80\% of the MO crystallized in reduced cases, and after only 40\% in the oxidized case (see Figure \ref{fig:Fe3+_fO2}b, when ${\rm IW}\sim1$). Hence, the evolution of the MO toward oxidized conditions accentuates the favorable effect of $f_{{\rm O}_2}$ on C retention. Overall, the final quantity of C in the mantle is multiplied by 2 upon switching from the reduced to the oxidized case, while the retention of H increases by at most 7\%.

While we consider an RCMF value of 0.4, the volume fraction of trapped melt in the cumulates may be lower, either because of an intrinsically lower value of the melt fraction at which the dynamical regime transition occurs, or because of melt extraction from the cumulates pile by percolation \citep{HierMajumderHirschmann2017}. The amount of volatiles retained in the solid mantle scales with the the amount of trapped melt, i.e. the RCMF. Adopting a lower value of the RCMF leads to enhanced fractionation (tending towards the fractional crystallization case as the RCMF becomes vanishingly small), and thus strengthens the increase in $f_{{\rm O}_2}$ as the MO crystallizes. Although the increase of $f_{{\rm O}_2}$ at cumulates crystallization associated with a lower RCMF can affect the solubility of volatiles, based on our results, we expect the MO cumulates retention to be roughly linear with RCMF value.
We find that decreasing the RCMF to 0.2 consistently halves the amount of C in the mantle throughout the parameter space, while the amount of H retained is multiplied by $\sim0.7$, without affecting the trends described above. The non-linear behaviour of H retention is attributable to the non-linear solubility of water, which is largely setting H solubility.

These results confirm the low efficiency of H outgassing associated with MO crystallization \citep{Miyazaki2022b,Bower2022}, and suggest that, beyond Earth, terrestrial planets and exoplanets which accreted from volatile-rich material start from a wet interior post-MO stage. This can have drastic consequences on their long-term mantle dynamics, as water significantly decreases both the melting temperatures, and the viscosity of silicate rocks. If a plate tectonic regime starts directly after MO crystallization, this could promote convection and thus result in a high plate velocity, accelerating the drawdown of atmospheric CO$_2$ and bringing the planet out of the runaway greenhouse state and toward habitable conditions \citep{Miyazaki2022a,Miyazaki2022b}. If the planet is in the stagnant lid regime, a wet mantle would lead to abundant melting. However, the latter may not necessarily translate into efficient outgassing if the atmosphere is thick enough to increase the solubility of water in basatic melt \citep{GaillardScaillet2014,Tosi2017,Godolt2019}. Thus the MO-outgassed atmosphere could prevent further outgassing of stagnant lid planets.

\subsection{Graphite saturation}
\label{subsec:graphite_saturation}
The volatile content of the system can be further altered by processes that have been neglected in the model.
The low $f_{{\rm O}_2}$ reached in the depth of the MO in the reduced cases (down to ${\rm IW}-4$, see Figure \ref{fig:initial_redox_profiles}a) decreases the CO and CO$_2$ concentrations at graphite saturation \citep{Yoshioka2019,EguchiDasgupta2018}, making precipitation of C possible. If the concentration of either (or both) species, as predicted by our model, increases above saturation, graphite (or diamond) would be produced to buffer the MO concentration at this value. Depending on the buoyancy of the precipitate, it could either settle at the bottom and leave the system (similar to the silicate crystals in the fractional crystallization case), remain entrained in the MO (as the silicate crystals in the equilibrium crystallization case), or accumulate at the surface and form a floating graphite layer (in which case the atmosphere redox state might be controlled by the CCO buffer \citep{KepplerGolabeck2019}). However, we notice that all our cases also contain H, which could form CH species (particularly at high pressure, \citep{Chi2014,Li2015,Gaillard2022a}) whose solubility could offset C saturation. \citet{Gaillard2022a} suggests that above 35 GPa (corresponding to the most reducing domain of the MO), C-content at C saturation should be above 100 ppm in H-bearing silicate melts, which is largely above the C concentration in our most C-rich case. However, it is important to keep in mind that this result relies on parametrizations calibrated for much lower temperatures and pressures than those met in a deep MO.

\subsection{Atmospheric escape}
\label{subsec:atm_escape}
%
Another process that has not been accounted for and could yet affect the H budget is atmospheric erosion. In the atmosphere, H$_2$, the lightest species, is particularly vulnerable to thermal escape, triggered by the enhanced XUV flux of the young sun. If H$_2$ is abundant in the exosphere (the region of the atmosphere from where it can readily escape), escape is limited by the energy provided by the XUV flux (energy-limited escape) while if diffusion of H$_2$ into the atmosphere is slow (and/or the XUV flux is low), H$_2$ availability in the exosphere becomes the bottleneck setting the escape rate (diffusion-limited escape). The energy-limited case provides a theroretical maximum for the escape flux, when all the incomming solar XUV is used for H$_2$ escape. Following \citet{Watson1981}, the H$_2$ mass loss rate in the energy-limited case is:
\begin{equation}
    \dot{M}_{{\rm H}_2}=\frac{R_{\rm sfc}\pi R_{\rm XUV}^2\epsilon F_{\rm XUV}(t)}{\mathcal{G}M_p},
    \label{eq:energy-limited_escape}
\end{equation}
where $R_{\rm XUV}$ is the radius at which XUV energy is effectively deposited, $\epsilon$ an XUV heating efficiency factor, $F_{\rm XUV}(t)$ the (time-dependent) received XUV flux, $\mathcal{G}$ the gravitational constant and $M_p$ the mass of the Earth. The time-dependence in $F_{\rm XUV}$ accounts for the increased XUV activity of the young sun. Following \citet{Katyal2020}, we calculate $\dot{M}_{{\rm H}_2}$ taking $R_{\rm XUV}=R_p$, $\epsilon=0.5$ and $F_{\rm XUV}(t)=5\times(4.5/t)^{1.24}\times10^{-3}$ where $t$ is the time elapsed after CAIs formation (i.e. the age of the sun) in Gyr, and $F_{\rm XUV}$ is expressed in W/m$^2$. Evaluating Eq. \ref{eq:energy-limited_escape} for a 100 Myr-old sun, we find an energy-limited H$_2$ mass flux of $6.32\times10^5$ kg/s. 
Using this value, we can calculate {\it a posteriori} the maximum mass of H lost over the lifetime of the MO for all cases in our parameter space. We find that it represents at most 4\% of the initial MO content, which is too small to have a significant impact on the chemical and/or thermal evolution of the atmosphere and the MO.

We note, however, that this effect might be important for exoplanets orbiting M-dwarfs, which have a much larger XUV flux during their early stages \citep{LugerBarnes2015,Barth2021}. In such cases, escape of H results in a net oxidation of the MO+atmosphere system \citep{Wordsworth2018}, which would alter the equilibrium between ferrous and ferric iron oxides. Another case where atmospheric escape could be relevant is when the incomming stellar radiation is higher than the OLR limit imposed by water condensation. In this case the planetary heat balance cancels out, and the MO no longer crystallizes. H$_2$ escape is then the main mechanism to alter the atmosphere, and increase the OLR until mantle eventually solidifies. This mechanism is sensitive to the stellar flux (hence the orbital distance), and for sun-like stars, a threshold in the semi-major axis has been suggested to be located between Venus's and Earth's orbits \citep{Hamano2013,Salvador2017}. A similar mechanism operating around other stars could allow for observable, long-lived MO exoplanets.

\subsection{Turbulent convection scaling}
\label{subsec:hard_vs_soft}

The MO lifetimes obtained in Section \ref{sec:results} are too short the for atmospheric loss to be significant. However, \citet{Nikolaou2019} found that adopting the hard turbulence (Equation \ref{eq:hard_turbulence}) rather than the soft turbulence (Equation \ref{eq:soft_turbulence}) scaling for the convective heat flux results in an increased MO lifetime.
The two essential differences between these two scalings are the dependence of the heat flux on the Rayleigh number and the Prandtl number. In particular, for a given Ra, the hard turbulence scaling yields a lower heat flux than the soft one as long as ${\rm Ra}\geqslant1.7\times10^8\times{\rm Pr}^3$, which is always the case in our fractional crystallization simulations, and holds until 97\% crystallization in the equilibrium cases. Conversely, sustaining the same heat flux requires a higher Ra in the hard turbulence scaling.

For the same set of parameters (volatiles budget, initial redox state and crystallization scenario), at a given stage of MO crystallization (i.e. a given $T_{\rm pot}$), all quantities entering the definition of Ra are fixed, except $T_{\rm sfc}$, while the Prandtl number is completely determined. Thus, the lower heat flux obtained in the hard turbulence scaling can be offset by decreasing $T_{\rm sfc}$ (i.e. increasing $\Delta T$, the temperature drop through the thermal boundary layer). In turn, decreasing $T_{\rm sfc}$ decreases the temperature in the atmosphere (where the composition is also fully characterized by $T_{\rm pot}$), which decreases the radiative flux. This latter effect induces a negative feedback, and the convective and radiative fluxes are balanced at a slightly lower value than in the soft turbulence scaling, although the Rayleigh number is higher.

\begin{figure}[h!]
    \centering
    \includegraphics[width=\textwidth]{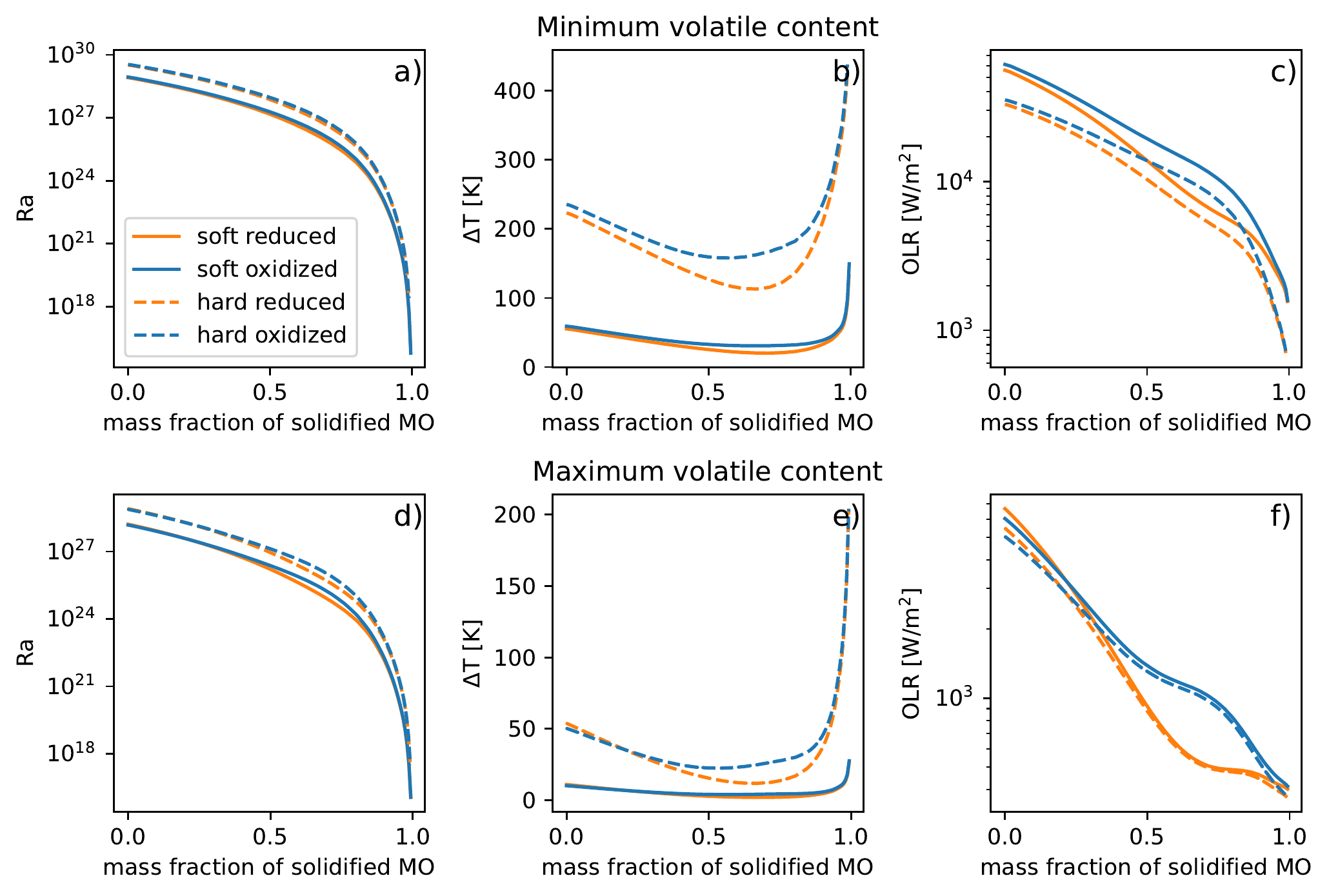}
    \caption{For the ``soft'' (blue curves) and ``hard'' (orange curves) turbulence scalings, evolution of (a) and (d) the Rayleigh number, (b) and (e) the temperature drop accross the thermal boundary layer, (c) and (f)  the OLR. Panels a-c correspond to the most volatil-depleted case, and panels d-f to the most volatile-rich case of the parameter space. In all panels, the $x$-axis is the mass fraction fraction of solidified MO (as in Figures \ref{fig:Fe3+_fO2} and \ref{fig:pressures_solubilities}).}
    \label{fig:hard_vs_soft}
\end{figure}

This effect is illustrated for a MO undergoing equilibrium crystallization with 1 EO H in the BSE and C/H=0.5, in Figure \ref{fig:hard_vs_soft}a-c. To a given position on the $x$-axis corresponds a given $T_{\rm pot}$. For the convective flux to balance the radiative flux, a higher Ra is needed when the hard turbulence scaling is adopted (dashed curves, Figure \ref{fig:hard_vs_soft}a), which is provided by a larger $\Delta T$ (Figure \ref{fig:hard_vs_soft}b). The resulting cooling flux of the MO is lower in the hard turbulence case during the whole MO crystallization (Figure \ref{fig:hard_vs_soft}c). 

We anticipate this effect to lose importance as the heat flux (and accordingly the Ra) diminishes, until vanishing when ${\rm Ra}\sim1.7\times10^8{\rm Pr}^3$. We confirm it by sweeping through our parameter space: we show that the prolongation of the MO lifetime due to adopting the hard turbulence scaling is important when the volatile budget is low (i.e. the radiative flux is generally high), and lessens as the volatile budget increases, yielding a thicker atmosphere. Figure \ref{fig:hard_vs_soft}d-f is similar to Figure \ref{fig:hard_vs_soft}a-c, but for a case having the maximum volatiles budget. While there still is a clear difference in the Ra (Figure \ref{fig:hard_vs_soft}d) and associated $\Delta T$ (Figure \ref{fig:hard_vs_soft}e) between hard and soft turbulence scalings, the resulting MO cooling fluxes are much more similar (Figure \ref{fig:hard_vs_soft}f).

Overall, for equilibrium crystallization cases (where the atmosphere is in general thinner), adopting the hard turbulence scaling yields an increase in MO lifetime from 73.5\% for the most volatile-depleted case (represented in Figure \ref{fig:hard_vs_soft}a-c), to 10.8\% for the most volatile-rich case (represented in Figure \ref{fig:hard_vs_soft}d-f). For fractional crystallization, where outgassing is much more extensive, this increase is limited between 8.3\% (most volatile-depleted case) and 1.3\% (most volatile-rich case). All these results are for initially oxidized cases; the lower heat flux resulting from initially reduced cases described in Section \ref{subsec:thermal_evolution} yields slightly lower MO lifetime prolongations. While the effect of using the hard turbulence scaling is more important for thinner atmospheres, which are more vulnerable to atmospheric escape, the prolongations are still not large enough for the energy-limited H-escape to alter significantly the planetary volatile budget during the MO lifetime.

Finally, it should be added that the dependence of the heat flux on the planetary rotation has also been omitted. According to \citet{Solomatov2015}, it becomes significant when Ra drops below $10^{19}$, which only occurs during the few last wt\% crystallization of the MO (see solid curves on Figures \ref{fig:hard_vs_soft}a and d). Thus, this effect will likely not affect the bulk of the MO solidification, although we notice that the last stages of MO crystallization are the longest. As the rotation of the Earth after the Moon-forming impact is unknown, we defer the investigation of the effect of the rotation to future studies.

\section{Concluding remarks}
By tracking the evolution of $f_{{\rm O}_2}$ in the MO during its crystallization and accounting for its influence on atmospheric speciation, we investigated the couplings between redox chemistry and volatile solubility, and their impact on MO outgassing and greenhouse effect. We reached the following conclusions:
\begin{enumerate}
    \item The incompatibility of Fe$^{3+}$ in the minerals in the crystallization sequence of the terrestrial MO results in its progressive enrichment, and in turn, in the increase of the $f_{{\rm O}_2}$ throughout the crystallization.
    \item Depending on its initial redox state (set by the effective conditions of core-mantle equilibration), the MO can either start from a mildly reduced state ($\Delta{\rm IW}=-2$ at the surface) and reach a mildly to highly oxidized one ($\Delta{\rm IW}=3$ to $\Delta{\rm IW}=8$), or start slightly more oxidized ($\Delta{\rm IW}=0$) and reach higher values.
    \item The atmosphere in equilibrium with the MO at the final stages of crystallization is consistently oxidized, dominated by H$_2$O and CO$_2$.
    \item The bulk H budget has a first-order influence on the lifetime of the MO, yielding a crystallization duration between 67 kyr and 1.53 Myr over the parameter space investigated here. The C budget and crystallization scenario have a second-order impact. The enhanced fractionation resulting from fractional crystallization yields a more extensive outgassing, as well as a more oxidized final state.
    \item Contrarily to previous studies, we find that initially reduced MOs have a longer lifetime than initially oxidized ones, in spite of the stronger greenhouse effect of oxidized species. This is due to the steeper lapse rate in reduced atmosphere, which yields a lower OLR and thus slower cooling.
    \item The retention of volatiles in melts trapped in the solid cumulates can amountup to 93\%of the bulk volatile budget. This mass of volatiles constitutes a reservoir than will possibly undergo further redox speciation influenced by solid-mantle buffers, and can alter the chemistry of the post MO atmosphere if released at the surface.
    \item The most volatile-depleted cases are most sensitive to the scaling used for the turbulent heat flux through the MO: adopting the hard turbulence scaling rather than the soft one yields an increase in the MO lifetime of up to 75\% for these cases.
\end{enumerate}

While these conclusions have been drawn in the context of the terrestrial MO formed after the Moon-forming giant impact, the processes investigated apply to early evolution of all terrestrial planets. Some parameters, such as initial volatile contents can vary significantly depending on the accretion history, but the trend of increasing oxidation state with ongoing MO crystallization and its associated consequences on MO-outgassed atmosphere is a general feature of rocky planets and exoplanets MO.

\subsection{Acknowledgments}
The authors thank two anonymous reviewers for their constructive comments which helped improve the quality of this study. M. M. thanks D. Bower for insightful discussions at an early stage of this project, J. Deng for sharing the Python routines used to calculate the pressure-dependence of the $f_{{\rm O}_2}$, and P. Mollière for his help with using petitRADTRANS.

\subsection{Data Availability}
The Python code and input files used to run the simulations and generate the figures presented in this study are freely available at: \href{https://bitbucket.org/MaximeMaurice/moai/src/master/}{https://bitbucket.org/MaximeMaurice/moai/src/Maurice\_etal\_2023/}, and is thoroughly documented at: \href{https://planetomoai.readthedocs.io/en/latest/index.html}{https://planetomoai.readthedocs.io/en/latest/index.html}

\appendix
\renewcommand{\thefigure}{A\arabic{figure}}
\setcounter{figure}{0}

\section{Extended chemical model}
\subsection{Iron oxides redox equilibrium}
The equilibrium between ferrous- and ferric-iron oxides is calculated using the parametrization from \citet{Hirschmann2022}:
\begin{eqnarray}
    \log \left(\frac{X_{{\rm FeO}_{1.5}}^{\rm mol}}{X_{{\rm FeO}}^{\rm mol}}\right) & =a\log f_{O_2}+b+\frac{c}{T}-\frac{\Delta C_p}{R\ln(10)}\left[1-\frac{T_0}{T}-\ln\left(\frac{T}{T_0}\right)\right]-\frac{\int_{P_0}^P\Delta Vdp}{RT\ln(10)} \nonumber \\
    &+\frac{1}{T}[\gamma_1X_{{\rm SiO}_2}^{\rm mol}+\gamma_2X_{{\rm TiO}_2}^{\rm mol}+\gamma_3X_{\rm MgO}^{\rm mol}+\gamma_4X_{\rm CaO}^{\rm mol}+\gamma_5X_{{\rm NaO}_{0.5}}^{\rm mol} \nonumber \\
    &+\gamma_6X_{{\rm KO}_{0.5}}^{\rm mol}+\gamma_7X_{{\rm PO}_{2.5}}^{\rm mol}+\gamma_8X_{{\rm SiO}_2}^{\rm mol}X_{{\rm AlO}_{1.5}}^{\rm mol}+\gamma_9X_{{\rm SiO}_2}^{\rm mol}X_{\rm MgO}^{\rm mol}]
    \label{eq:fO2_Hirschmann2022}
\end{eqnarray}
where $X_i^{\rm mol}$ is the molar concentrations of oxide $i$, $f_{O_2}$ the oxygen fugacity, $R$ the gas constant, $P_0=$ $10^5$ Pa the reference state pressure, and $a$, $b$, $c$, $\Delta C_p$, $T_0$ and $\gamma_{1,...,9}$ are fitted parameters given in Table 2 of \citet{Hirschmann2022}. The composition in all oxides is taken from the BSE composition of \citet{BSE}. Notice that for iron oxide, only the initial value of FeO+FeO$_{1.5}$ corresponds to \citet{BSE}.

In order to determine the initial ferric-to-total iron ratio, we evaluate Equation \ref{eq:fO2_Hirschmann2022} at the $P$-$T$-conditions met at the surface and at the bottom of the magma ocean in the ``reduced'' and ``oxidized'' cases respectively (see Figure \ref{fig:initial_redox_profiles}), and at $\Delta{\rm IW}=-2$ at the corresponding $P$-$T$ conditions, where IW$(P,T)$ is given by \citet{Hirschmann2021}.

Subsequently, to calculate the evolution of the $f_{{\rm O}_2}$ in the atmosphere, we use eq \ref{eq:fO2_Hirschmann2022} to compute $f_{{\rm O}_2}$ at the $P$-$T$-conditions met at the surface of the magma ocean, with $X_{{\rm FeO}_{1.5}}^{\rm mol}/X_{{\rm FeO}}^{\rm mol}$ given by the differential partitioning of each iron oxide (see Section \ref{apx:part_coefs}).

\subsection{Initial redox state}
\label{apx:initial_redox_state}
\begin{figure}[h!]
    \centering
    \includegraphics[width=\textwidth]{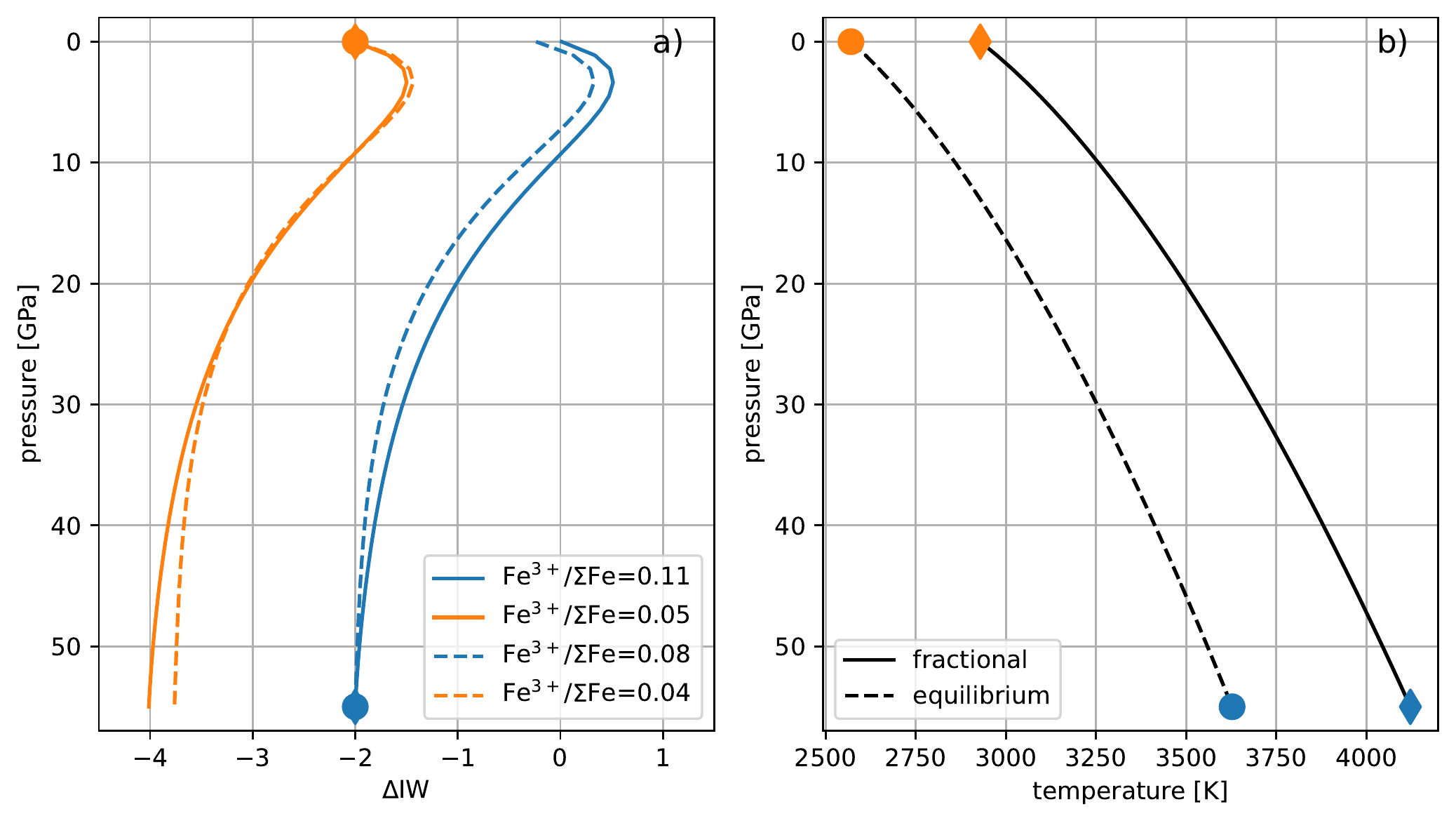}
    \caption{Initial profiles of the $f_{{\rm O}_2}$ (expressed as $\Delta{\rm IW}$) (a) and the temperature (b) for all distinct cases. The linestyles in both panels and the colors in panel a) correspond to those used in Figure \ref{fig:Fe3+_fO2}: orange for reduced cases and blue for oxidized ones, solid for fractional crystallization and dashed for equilibrium crystallization. The markers indicate the $P$-$T$-$f_{{\rm O}_2}$ conditions of core-mantle equilibrium ($\Delta{\rm IW}=-2$): orange for reduced cases (equilibration at the surface) and blue for oxidized ones (equilibration at 55 GPa), diamond for fractional crystallization and circles for equilibrium crystallization (notice that markers of the same color overlap in panel a).}
    \label{fig:initial_redox_profiles}
\end{figure}

The initial $f_{{\rm O}_2}$ profile of the oxidized and reduced fractional crystallization cases are represented in Figure \ref{fig:initial_redox_profiles}. The respective values of the initial ${\rm Fe}^{3+}/\Sigma{\rm Fe}$ are calculated at the effective conditions of core-mantle equilibration (represented by the markers in both panels), to match $\Delta{\rm IW}=-2$. The difference in $f_{{\rm O}_2}$ profiles between the fractional and equilibrium crystallization cases sharing the same core-mantle equilibration depth (i.e. between curves of the same color) is due to the colder initial temperature profile in the latter case, which intercepts the liquidus at 55 GPa, while the former intercepts the RCMF.

\subsection{Ferric iron oxide partition coefficients}
\label{apx:part_coefs}
We calculate the partition coefficient of ferric iron oxide at each crystallization step as the mass-weighted average of the mineral-melt partition coefficient of ferric iron oxide in each mineral present in the composition of the crystallized cumulates layer. We first calculate a pressure profile of ferric iron oxide partition coefficient for the crystallization sequence of \citet{ElkinsTanton2008} (for a 2000-km-deep MO, truncated at 55 GPa). In the equilibrium crystallization scenario, the effective partition coefficient of ferric iron oxide is equal to this profile averaged over all pressures between the bottom of the MO and the pressure at which the adiabat intercepts the liquidus. In the fractional crystallization scenario, the partition coefficient of ferric iron oxide is equal to the value on this profile at the pressure of the bottom of the MO. Individual mineral-melt partition coefficients for ferric iron oxide are listed in Table \ref{tab:part_coefs}.

\begin{table}[h!]
    \centering
    \begin{tabular}{l l l}
        \textbf{Mineral} & \textbf{Value} & \textbf{source} \\ \hline
        Olivine & 0.05 & \citet{MallmannONeill2009} \\
        Orthopyroxene & 0.18 & \citet{MallmannONeill2009} \\
        Clinopyroxene & 0.44 & \citet{MallmannONeill2009} \\
        Garnet & 0.37 & \citet{CanilONeill1996} \\
        Wadsleyite & 0.59 & \citet{FrostMcCammon2009} \\
        Ringwoodite & 0.2 & \citet{Deon2011} \\
        Majorite & 0.63 & \citet{CanilONeill1996} \\
        Bridgmanite & 0.75 & \citet{Boujibar2016} \\
        Periclase & 0 & \citet{Otsuka2013} \\
        Plagioclase & 3.8 & \citet{LundgaardTegner2004} \\ \hline
    \end{tabular}
    \caption{Mineral-melt partition coefficients for ferric-iron oxide.}
    \label{tab:part_coefs}
\end{table}

\subsection{volatiles redox equilibria}
\label{apx:volatile_redox}
Equilibirum constants for the volatile redox reactions are calculated form the Gibbs-free energy of reaction ($\Delta_fG_0$), itself calculated from the Gibbs-free energy of each of the reactants and products ($\Delta_rG_X$):
\begin{eqnarray}
    K_{\rm H}(T)&=\exp\left(\frac{\Delta_fG_{0,{\rm H}}(T)}{RT}\right)\nonumber \\
                &=\exp\left(\frac{-2\Delta_rG_{{\rm H}_2{\rm O}}(T)}{RT}\right), \\
    K_{\rm C}(T)&=\exp\left(\frac{\Delta_fG_{0,{\rm C}}(T)}{RT}\right)\nonumber \\
                &=\exp\left(\frac{-2\Delta_rG_{{\rm CO}_2}(T)+2\Delta_rG_{{\rm CO}}(T)}{RT}\right),
\end{eqnarray}
where $T$ is the temperature and $R$ the gas constant. The values of the species formation Gibbs-free energy are interpolated at the desired temperature from the NIST tables (\href{https://webbook.nist.gov/}{https://webbook.nist.gov/}).

\subsection{Volatiles solubilities}

The volatile solubility laws are given in Equation \ref{eq:Henry_law}. The species-specific $\alpha$ and $\beta$ coefficients ar given in Table \ref{tab:solubility_coefs}.

\begin{table}[h!]
    \centering
    \begin{tabular}{l l l l l}
        {\bf species} & $\alpha^a$ & $\beta$ & {\bf reference} \\
        \hline
        H$_2$ & $5\times10^{-6}$ & 1 & \citet{Hirschmann2016} \\
        H$_2$O & $1.7$ & 1/2 & \citet{Bower2022} \\
        CO & $5.5\times10^{-7}$ & 1 & \citet{Hirschmann2016} \\
        CO$_2$ & $1.6\times10^{-6}$ & 1 & \citet{Hirschmann2016} \\ 
    \end{tabular}
    \caption{Solubility law coefficients for the different gaseous species. $^a$: $\alpha$ is expressed in ppm/Pa$^{1/\beta}$. Notice that $\alpha$ values of H$_2$, CO and CO$_2$ are given for element solubility in \citet{Hirschmann2016}, and are here corrected for species solubility.}
    \label{tab:solubility_coefs}
\end{table}

\subsection{mass conservation}
\label{apx:mass_cons}
In both crystallization scenarios, the potential temperature entirely defines the advancement in crystallization and thus the mass of the chemical system $M_{\rm syst}$. Hence, we can express the mass change $dM_{\rm syst}$ as a function of $dT_{\rm pot}$: $dM_{\rm syst}=dM_{\rm syst}/dT_{\rm pot}\times dT_{\rm pot}$. We obtain $dT_{\rm pot}$ from the thermal evolution model (Eq. \ref{eq:heat_cons_planet}). How $dM_{\rm syst}/dT_{\rm pot}$ is obtained in either crystallization scenario is described below.

In the equilibrium crystallization scenario, the bottom boundary of the chemical system is defined by the intercept between the adiabat and the RCMF temperature, we thus have: $M_{\rm syst}=4/3\pi(R_p^3-R_{\rm RCMF}^3(T_{\rm pot}))$, where $R_{\rm RCMF}(T_{\rm pot})$ is the radius at which the adiabat (having potential temperature value $T_{\rm pot}$) intercepts the RCMF. Taking the derivative of this expression yields:
\begin{equation}
    \frac{dM_{\rm syst}}{dT_{\rm pot}}=-4\pi R_{\rm RCMF}^2(T_{\rm pot})\frac{dR_{\rm RCMF}}{dT_{\rm pot}},
    \label{eq:dMsyst_equi}
\end{equation}
which can be calculated from the shape of the melting curves and the equations of state, by mapping $R_{\rm RCMF}$ onto $T_{\rm pot}$ and taking the derivative of the function thus obtained.

In the fractional crystallization scenario, the decrease in the chemical system's mass is exactly equal to the decrease in melt mass (i.e. the increment in crystals mass) due to crystallization, which is calculated by integrating the melt fraction , and thus:
\begin{equation}
    \frac{dM_{\rm syst}}{dT_{\rm pot}}=4\pi\int_{R_{\rm MO}(T_{\rm pot})}^{R_p}\frac{\partial\phi(r)}{\partial T_{\rm pot}}r^2dr,
    \label{eq:dM_syst_frac}
\end{equation}
where $R_{\rm MO}(T_{\rm pot})$ is the bottom of the MO when the potential temperature is $T_{\rm pot}$. When evaluating Eq. \ref{eq:dM_syst_frac}, $\phi$ is obtained from Eq. \ref{eq:melt_fraction}, even though it does not represent the actual profile of the melt fraction in the fractional crystallization scenario since crystals are assumed to settle (the melt fraction profile is just a step function). Actually, Eq. \ref{eq:melt_fraction} does represent the melt fraction profile in the MO for the vanishingly small (virtual) instant between cooling of the adiabat and settling of the crystals, as represented in Figure \ref{fig:crystallization_scenarios}.

\section{Extended thermal model}

\subsection{Secular cooling}
\label{apx:secular_cooling}
The secular cooling term in Equation \ref{eq:dTpot_dt} is calculated as $\Delta E_{\rm th}/\Delta T_{\rm pot}$, where $\Delta E_{\rm th}$ is the difference between the thermal energy of the initial (hot) and final (cold) states. Here we present the calculation of $\Delta E_{\rm th}$ based on the fractional crystallization case, which differs from $\Delta E_{\rm th}$ in the equilibrium crystallizatoin case, but the ratio $\Delta E_{\rm th}/\Delta T_{\rm pot}$ is assumed to be the same. In the initial state, the whole planet follows a single adiabat whose slope depends on the equation of state of the ambient material (liquid MgSiO$_3$ in the mantle \citep{deKokerStixrude2013}, liquid iron in the core \citep{SaxenaEriksson2015}), given by the initial potential temperature ($T_{\rm pot}^i=$2939 K, corresponding to a 55 GPa-deep MO). In final state, the temperature in the mantle is given by the adiabat from a potential temperature is equal to the surface temperature of the solidus ($T_{\rm pot}^f=$1362 K \citep{Fiquet2010}, hence $\Delta T_{\rm pot}=1577$ K), and the temperature in the core follows an adiabat from the estimated present-day CMB temperature (4400 K \citep{Andrault2016}). This yields a total thermal energy difference of $1.31\times10^{31}$ J, and in turn, a secular cooling term of $8.34\times10^{27}$ J/K. This value, calculated with initial and final states corresponding to the fractional crystallization, apply to the equilibrium crystallization as well, since $dE_{\rm th}/dT_{\rm pot}$ is independent from the potential temperature span covered.
\begin{figure}
    \centering
    \includegraphics[width=13cm]{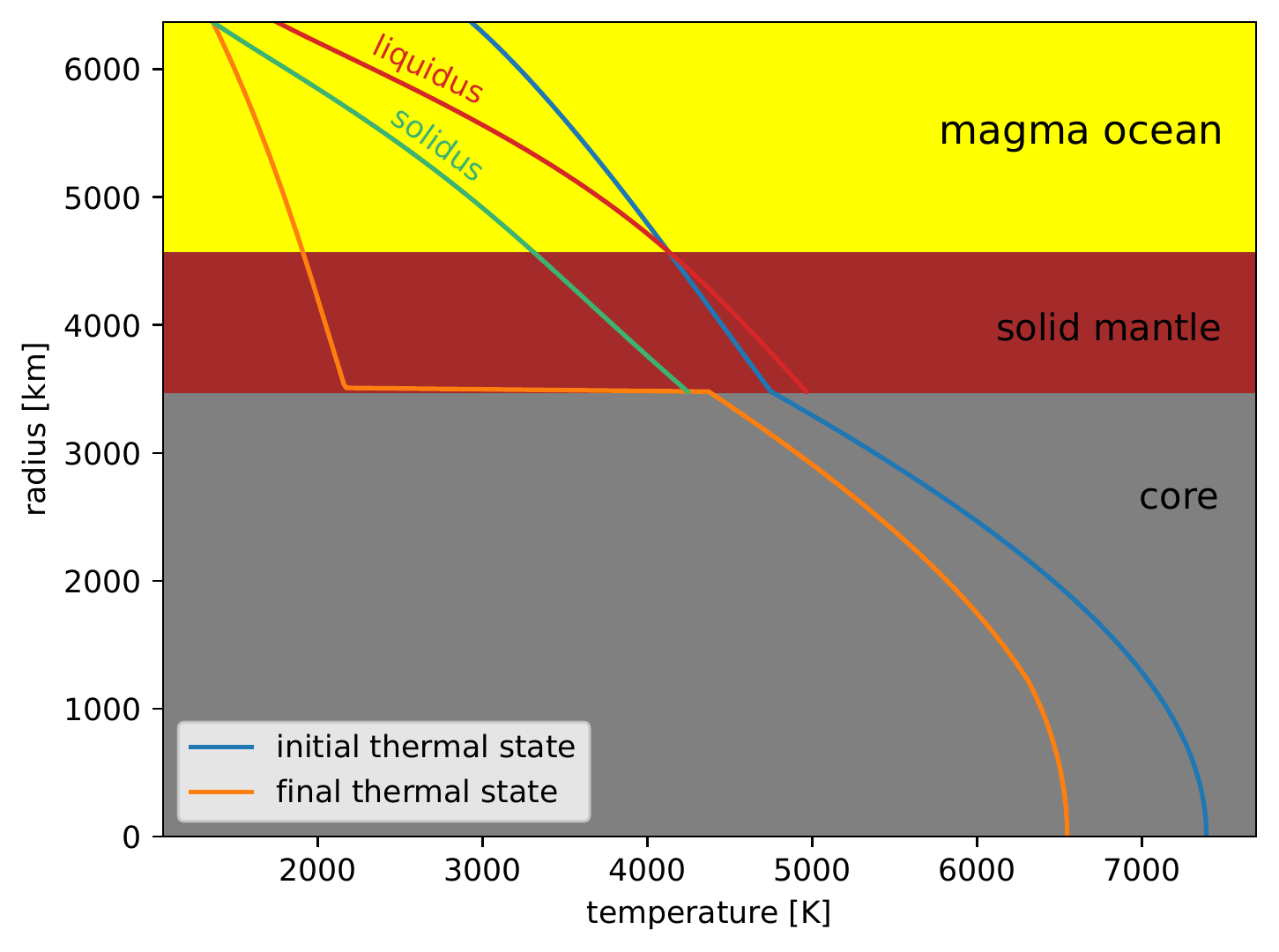}
    \caption{Initial (blue) and final (orange) thermal states from which the secular cooling term is calculated.}
    \label{fig:thermal_states}
\end{figure}

\subsection{MO viscosity}
\label{ap:MO_visc}
The viscosity of silicate melts depends on $T$ following a Vogel-Fulcher-Tammann law (as in \citet{Nikolaou2019}):
\begin{equation}
    \eta(T)=A_K\exp{\left(\frac{B_K}{T-C_K}\right)},
    \label{eq:VFT}
\end{equation}
with $A_K=2.4\times10^{-4}$ Pa s, $B_K=4600$ K and $C_K=1000$ K.

We do not consider the dependence of $\eta$ on the water content in the melt. While we find that this content can increase by one order of magnitude during MO crystallization, reaching the order of $\sim2$ wt.\% (thus possibly decreasing $\eta$ by about one order of magnitude), this effect is small compared to the general increase of $\eta$ by several orders of magnitude due to temperature decrease as the MO crystallizes (see Figure 2a of \citet{Nikolaou2019}). Furthermore, no model for the influence of water on silicate melt viscosity has been calibrated on a peridotitic composition, to our knowledge. \citet{Nikolaou2019} used a parametrization calibrated on basanitic composition (with a modified pre-factor), which differs significantly with the BSE composition that we consider in this study.

In the equilibrium crystallization case, the viscosity in the MO is altered by the presence of crystals. The effective viscosity (used to compute the convective heat flux in the MO) depends on $\phi$ and reads (following \citet{Roscoe1952}):
\begin{equation}
    \eta_{\rm eff}(T,\phi)=\frac{\eta(T)}{\left(1-\frac{1-\phi}{1-\phi_{\rm RCMF}}\right)^{2.5}},
\end{equation}
where $\eta(T)$ is given Equation \ref{eq:VFT}.

\subsection{Water saturation pressure}
\label{apx:p_sat}
In the moist troposphere, the partial pressure of water ($p_{{\rm H}_2{\rm O}}$) is equal to its saturation pressure at ambient temperature, given by \citep{Pierrehumbert2010}:
\begin{equation}
    p_{{\rm H}_2{\rm O}}(T)=p_{\rm ref}\exp{\left(-\frac{L_{{\rm H}_2{\rm O}}}{R}\left(\frac{1}{T}-\frac{1}{T_{\rm ref}}\right)\right)},
    \label{eq:psat}
\end{equation}
where $p_{\rm ref}$ and $T_{\rm ref}$ correspond to a reference point on the saturation curve of water, here taken to be the triple point (see Table \ref{tab:param3}). Equation \ref{eq:psat} derives from the Clausius-Clapeyron equation with the assumption that the latent heat is constant.

\section{Sensitivity of the MO lifetime}

\label{apx:duration_sensitivity}
Many factors affect the lifetime of the terrestrial MO \citep{Nikolaou2019,Monteux2016}. In this study, simplifications were made for the sake of interpretability of the model results. We discuss the anticipated consequences of accounting for the neglected processes in this section.

\subsection{Approximations in the secular cooling}
\label{apx:thermal_evolution_approx}
Our thermal evolution is simple and the secular cooling term in particular lacks complexity. It aims of providing an upper bound for the MO lifetime, which can then be used as a conservative estimate e.g. to address the effect of atmospheric escape while the MO is extent. The secular cooling term, as it is defined (Section \ref{apx:secular_cooling}) and used (Section \ref{subsubsec:heat_cons}, has three main sources of inaccuracies or simplifications. First, $dE_{\rm th}/dT_{\rm pot}$ is probably not constant, as intrinsically time-dependent processes will be at play during MO crystallization, which have timescales comparable with the MO lifetime, for instance onset of solid-state convection in the solidified mantle \citep{Maurice2017,Ballmer2017}. As a result, separating the time and temperature dependence of the secular cooling term as implied in going from Eq. \ref{eq:heat_cons_planet} to Equation \ref{eq:dTpot_dt} might be inaccurate. However, resolving these processes is largely beyond the scope of this study, and taking $dE_{\rm th}/dT_{\rm pot}$ constant is a first order approach that we expect to yield correct order of magnitude results. Second, the initial thermal state of the post-giant-impact Earth is largely unconstrained. This thermal state is the result of the primordial accretion heating of the planet (subsequently dissipated by up to 150 Myr of thermal evolution prior to the Moon-forming impact) as well as internal heating provided by radioactive elements (operating over the same time span), and the heating from the impact itself. Here, the initial thermal state is entirely defined by the choice of a 55 GPa-deep MO as well as a temperature profile following one single adiabat. We make the asumption that such a thermal state is compatible with the anterior thermal evolution of the proto-Earth. Third, the final thermal state, at the end of the MO crystallization, is also the result of hypotheses, namely that the mantle and core follow each a single adiabat. The mantle might instead have a hotter temperature corresponding to the transient regime of solid-state convection onset (although solid-state convection might have time to be fully developped, in which case the temperature profile is likely adiabatic indeed). This assumption is in line with previous studies on terrestrial MO solidification \citep{Lebrun2013,Nikolaou2019}, easing the comparison between their results and ours. It is more likely that the core follows an adiabat, as inner core growth will not have initiated yet. However, the potential temperature of the core's adiabat (defined at the CMB) is uncertain. We evaluated the secular cooling term for a post-MO core temperature of 4400 as suggested by \citet{Andrault2016}. While a lower temperature is not excluded, we remark that the cooling of the core by 1000 K between the initial and final thermal states represents only about 10\% of the secular cooling, so an error on the core temperature of a few 100 K is unlikely to dramatically affect our results.

\subsection{Atmosphere evolution}
\label{apx:atm_evol}

In line with our assumption of total rain out of the moist troposphere (following \citet{Pierrehumbert2010}), we disregard cloud formation. Clouds are arguably the most complicated process influencing heat transfer through the atmosphere, as they can absorb infrared radiation (thus decrease the planet's cooling rate) as well as increase the albedo (thus increase the planet's cooling rate). The former effect can be accounted for in 1-D column atmosphere models like the one we use, but requires cloud microphysics parametrizations that add several new parameters (e.g. turbulence diffusion in the clouds, size distribution of droplets etc.). The latter effect requires 3-D circulation models that can resolve the cloud distribution. \citet{Pluriel2019} ran such simulations for CO$_2$-H$_2$O atmospheres and proposed a parametrization of the albedo as a function of the composition of the atmosphere and the surface temperature. However, here we also consider reduced species, which can alter the structure of the troposphere (in particular, the extension of the moist troposphere). Hence, the parametrization of \citet{Pluriel2019} may not be suitable for such atmospheres.
\bibliography{references.bib}



\end{document}